\documentclass[12pt]{amsart}
\usepackage[T1]{fontenc}
\usepackage{amssymb}
\usepackage{enumerate}
\usepackage{enumitem}
\usepackage[colorlinks=true, linkcolor=blue, urlcolor=blue, citecolor=blue, anchorcolor=blue, pdfborder={0 0 0}]{hyperref}
\usepackage{breqn}
\usepackage{color,colortbl}
\usepackage[dvipsnames,table,xcdraw]{xcolor}
\usepackage{cite}
\usepackage{amsmath, amssymb, amsthm, amsfonts}
\usepackage{mathtools}
\usepackage[top=45truemm, bottom=45truemm, left=35truemm, right=30truemm]{geometry}

\usepackage{nicefrac}
\usepackage{cancel}
\usepackage{float}
\usepackage{tabularx}
\usepackage{makecell}
\usepackage{array}
\usepackage{ragged2e}
\usepackage{graphicx}
\usepackage{caption}
\usepackage{subcaption}
\usepackage{listings}
\usepackage[frozencache,cachedir=.]{minted}
\usepackage[utf8]{inputenc}
\usepackage{textgreek}
\usepackage[symbol]{footmisc}
\usepackage{tikz}
\usepackage{orcidlink}
\usepackage{xcolor}
\usepackage[linesnumbered, ruled, vlined]{algorithm2e}
\makeatletter
\def\@settitle{\begin{center}%
  \baselineskip14\p@\relax
  \bfseries
  \uppercasenonmath\@title
  \@title
  \ifx\@subtitle\@empty\else
     \\[1ex]\uppercasenonmath\@subtitle
     \footnotesize\mdseries\@subtitle
  \fi
  \end{center}%
}
\def\subtitle#1{\gdef\@subtitle{#1}}
\def\@subtitle{}
\makeatother

\definecolor{bg}{rgb}{0.95,0.95,0.95}
\definecolor{LightGray}{RGB}{242,242,242}

\newcommand{\code}[1]{\colorbox{bg}{\texttt{#1}}}

\newcommand{\ket}[1]{\vert {#1} \rangle}

\newcolumntype{P}[1]{>{\RaggedRight\hspace{0pt}}p{#1}}
\newcolumntype{L}[1]{>{\raggedright\let\newline\\\arraybackslash\hspace{0pt}}m{#1}}
\newcolumntype{C}[1]{>{\centering\let\newline\\\arraybackslash\hspace{0pt}}m{#1}}
\newcolumntype{R}[1]{>{\raggedleft\let\newline\\\arraybackslash\hspace{0pt}}m{#1}}

\newtheorem{theorem}{Theorem}

\newtheorem{definition}{Definition}

\newtheorem{proposition}{Proposition}

\theoremstyle{definition}
\newtheorem{remark}{Remark}

\title{Quantum Error Propagation}

\author[Eldar Sultanow]{\orcidlink{0000-0001-5257-2236}\hspace{1mm}Eldar Sultanow}
\address{Eldar Sultanow\\Capgemini\\Bahnhofstraße 30\\Nuremberg, Germany}
\curraddr{}
\email{eldar.sultanow@capgemini.com}

\author[Fation Selimllari]{\orcidlink{0009-0008-3391-8404}\hspace{1mm}Fation Selimllari}
\address{Fation Selimllari\\Hochschule Coburg\\Friedrich-Streib-Str. 2\\Coburg, Germany}
\curraddr{}
\email{fation@selimllari.de}

\author[Siddhant Dutta]{\orcidlink{0009-0000-5120-7114}\hspace{1mm}Siddhant Dutta}
\address{Siddhant Dutta\\SVKM's Dwarkadas J. Sanghvi College of Engineering\\Bhaktivedanta Swami Rd\\Mumbai, India}
\curraddr{}
\email{siddhantdutta1@gmail.com}

\author[Barry D. Reese]{Barry D. Reese}
\address{Barry D. Reese\\Capgemini\\Olaf Palme Stasse\\Munich, Germany}
\curraddr{}
\email{barry.d.reese@gmail.com}

\author[Madjid Tehrani]{\orcidlink{0000-0002-4838-5865}\hspace{1mm}Madjid Tehrani}
\address{Madjid Tehrani\\The George Washington University\\School of Engineering and Applied Science\\Washington D.C., USA}
\curraddr{}
\email{madjid\_tehrani@gwu.edu}

\author[William J Buchanan]{\orcidlink{0000-0003-0809-3523}\hspace{1mm}William J Buchanan}
\address{William J Buchanan\\Edinburgh Napier University\\Edinburgh, UK}
\curraddr{}
\email{b.buchanan@napier.ac.uk}


\begin{document}

\begingroup
\let\MakeUppercase\relax
\clearpage\maketitle
\thispagestyle{empty}
\endgroup

\begin{abstract}
Data poisoning attacks on machine learning models \cite{yerlikaya2022data} aim to manipulate the data used for model training such that the trained model behaves in the attacker's favour. In classical models such as deep neural networks, large chains of dot products do indeed cause errors injected by an attacker to propagate or accumulate. But what about quantum models? We hypothesise that, in quantum machine learning, error propagation is limited for two reasons. The first is that data, which is encoded in quantum computing, is in terms of qubits that are confined to the Bloch sphere. Second, quantum information processing happens via the application of unitary operators, which preserve norms. Testing this hypothesis, we investigate how extensive error propagation and, thus, poisoning attacks affect quantum machine learning \cite{lu2020quantum}.

\end{abstract}

\section{Introduction}
\label{sec:introduction}

Quantum machine learning (QML) involves the application of machine learning programs with quantum algorithms. This quantum processing then uses qubits and quantum operations to enhance computational speed and data storage and where a hybrid system can be used with both classical and quantum computing \cite{biamonte2017quantum}. In general, QML \cite{abohashima2020classification} refers to the idea of utilizing quantum computing at various stages of the machine learning pipeline. Here, the expectation is that quantum advantages may lead to faster processing or better internal representations. However, our interest in this paper is in robustness, and the question of whether or not the fairly restricted algebraic structures that govern the world of quantum computing also offer advantages with respect to error mitigation. 

To begin with, we state our hypotheses and research question on error propagation in quantum machine learning. Answering our research question will require a formalization of the problem with respect to the behaviour of qubits. In order to describe error propagation geometrically, we investigate how the successive rotation of a vector on the Bloch Sphere \cite{goyal2016geometry} can be mathematically formalized.

\subsection{Hypotheses}
\label{sec:hypotheses}
Our two hypotheses are:
\begin{description}
\item[H1] The propagation of data errors -- discrepancies of the Euler angles \cite{tilma2002generalized} of qubits on the Bloch sphere \cite{wie2020two} -- is limited and therefore cannot grow indefinitely.
\item[H2] Periodically, errors increase and decrease back to zero. This periodic behaviour becomes more complicated the more and the stronger Euler angles are biased by an error.
\end{description}

\par\medskip
While these hypotheses may seem almost obvious, the following question is more difficult to answer:
\begin{description}
\item[Q] How can the periodic nature of error growth and decline be described mathematically (ideally in closed form) in order to allow for quantifying the effects of data poisoning in QML?
\end{description}
Observing the above fact, we developed a new hypothesis to evaluate whether this property will be preserved in a quantum circuit, such as the ZZFeatureMap used in QSVM, as described below to evaluate the robustness of quantum machine learning (QML) against poisoning attacks.
\[
H_0: \mu_{\text{QSVM-poisoned}} = \mu_{\text{SVM-poisoned}}
\]
\[
H_1: \mu_{\text{QSVM-poisoned}} > \mu_{\text{SVM-poisoned}}
\]

Where \(\mu_{\text{QSVM-poisoned}}\) and \(\mu_{\text{SVM-poisoned}}\) is the mean accuracies of QSVM and SVM under the poisoning scenario. We test if QSVM is significantly more robust.

To get a gentle introduction to algebraic structures underlying the error propagation theory, Appendix~\ref{appx:single-qubit-rotations} explores single-qubit operations with the aim of providing a foundational framework for understanding the mathematical principles governing quantum noise.

\section{Poisoning attacks of QML}

Recent advancements in cloud-based quantum machine learning (QML) have brought unprecedented opportunities and unique challenges to the field of adversarial machine learning. Classical Support Vector Machines (SVMs), as analyzed by Biggio et al. (2012)\cite{biggio2013poisoningattackssupportvector}, have long been a target for poisoning attacks, where malicious data compromises model integrity, causing significant error rate increases. Extending such concerns into the quantum domain, Franco et al. (2024)\cite{franco2024predominantaspectssecurityquantum} underscore vulnerabilities in QML, such as fault injection and quantum noise exploitation, and highlight the potential of adversarial training and quantum differential privacy as countermeasures. Wendlinger et al. (2024)\cite{wendlinger2024comparativeanalysisadversarialrobustness} further demonstrate the susceptibility of quantum models to adversarial perturbations, emphasizing the necessity for robust regularization techniques. Meanwhile, Kundu and Ghosh (2024)\cite{kundu2024securityconcernsquantummachine} detail the security risks in hybrid QMLaaS frameworks, exposing threats to data integrity and system stability, and propose encryption and trusted execution environments as defences. Li et al. (2024) \cite{li2024computablemodelindependentboundsadversarial} introduce lower bounds for adversarial error rates in QML, providing valuable benchmarks for model robustness. Notably, the QUID attack by Kundu and Ghosh (2024)\cite{kundu2024adversarialpoisoningattackquantum} achieves up to 90\% accuracy degradation in QML models under label-flipping data poisoning, illustrating the severe consequences of adversarial attacks.

Against this backdrop, the evaluation of Quantum Support Vector Machines (QSVMs) emerges as a critical research imperative. Yu and Zhou (2024) \cite{10688912} address adversarial resilience in power system applications with QaTSA but focus predominantly on tailored quantum circuits. Similarly, Reers and Maussner (2024) \cite{reers2024comparative} provide a broad comparative analysis of vulnerabilities in classical and quantum frameworks but leave gaps in QSVM-specific evaluations. Our study aims to fill this void, comparing the high-impact QUID attacks \cite{kundu2024adversarialpoisoningattackquantum} with the broader applicability of QSVMs in real-world scenarios. 

\section{Experiment Setup}
In this experiment, we utilized \textit{Qiskit version 1.3.1}, along with \textit{qiskit-machine-learning version 0.8.2} and \textit{qiskit-algorithms version 0.3.1} to implement and test quantum machine learning models and algorithmic simulations. The experiment was conducted on a \textit{Google Colab environment}, leveraging a \textit{Python 3 Google Compute Engine backend}. The computational resources included \textit{273.66 compute units}. The system provided \textit{51 GB of RAM} and a disk capacity of \textit{225.8 GB}.

All the code for reproducing this experiment is publicly available (see notebook \href{https://github.com/Sultanow/quantum/blob/main/qsvm_svm_poisoning/QSVM_SVM_Poisioning.ipynb}{QSVM\_SVM\_Poisioning.ipynb} on GitHub).

\section{Experiment Design for SVM vs. QSVM}
\subsection{Problem Definition}
We consider a binary classification task \cite{MathWorks2024}, where an adversary has a strong motivation to poison a QSVM algorithm deployed on cloud-based quantum machines:
\begin{itemize}
    \item \textbf{Class 1:} Cylinder
    \item \textbf{Class 2:} Cone
\end{itemize}

The radar cross-section (RCS) of an object is modelled as:
\[
\text{RCS}(r, h, \theta_{\text{az}}, \theta_{\text{el}}) = \frac{(r \cdot h)^2}{\lambda^2} \cdot \cos(\theta_{\text{az}}) \cdot \cos(\theta_{\text{el}}),
\]
with \(r,h\) the geometry parameters, \(\lambda\) the wavelength, and \(\theta_{\text{az}}, \theta_{\text{el}}\) the angles. We generate synthetic data with Gaussian noise in angles.

\subsubsection{Objective}
We aim to classify a given RCS profile as either a cylinder or a cone. We compare:
\begin{enumerate}
    \item Classical SVM: Unpoisoned vs. Poisoned.
    \item QSVM: Unpoisoned vs. Poisoned.
\end{enumerate}

We investigate whether the QSVM exhibits greater robustness to data poisoning than the classical SVM.
\subsubsection{assumption of adversarial access}
The attack assumes that the adversary has partial access to the training data and can inject a fraction of poisoned samples (\( \epsilon \)) while possessing knowledge of the model type (classical or quantum SVM), kernel functions, and data encoding schemes. For classical SVMs, the adversary exploits kernel-induced feature spaces, while for quantum SVMs (QSVMs), fidelity-based distances in the quantum Hilbert space are targeted. The attack presumes the availability of sufficient computational resources to craft adversarial samples that distort the decision boundary, leveraging the reproducibility of model training. Additionally, the system is assumed to operate in a fault-tolerant quantum environment, and no robust defences, such as adversarial training, anomaly detection, or encryption, are assumed to be in place during the attack.

\subsection{Methodology}
\begin{enumerate}
    \item \textbf{Data Generation:} Synthetic RCS data for both classes, creating training and test sets.
    \item \textbf{Feature Extraction:} Flatten RCS profiles and apply PCA to reduce dimensionality (e.g., to 10 components).
    \item \textbf{Normalization:} Normalize PCA-reduced features to [0,1], especially for QSVM.
    \item \textbf{Models:}
    \begin{itemize}
        \item \textbf{SVM:} Classical SVM with a polynomial kernel on reduced features.
        \item \textbf{QSVM:} Uses a fidelity-based quantum kernel (via a ZZFeatureMap and FidelityQuantumKernel).
    \end{itemize}
    \item \textbf{QUID-Inspired Poisoning:}  
    Instead of simple additive perturbations, we reassign labels of a subset of training samples based on a QUID-inspired strategy:
    \begin{itemize}
        \item For the SVM, compute distances in the kernel-induced feature space.
        \item For the QSVM, compute fidelity-based distances between quantum states.
    \end{itemize}
    The poisoned samples are assigned to the class that is, on average, the most distant.
    \item \textbf{Monte Carlo Runs:} Repeat the experiment multiple times(here, 30 times) to get distributions of accuracies.
    \item \textbf{Statistical Testing:} Perform a paired t-test on the difference in poisoned accuracies of QSVM and SVM. If \(p<0.05\), conclude QSVM is more robust.
\end{enumerate}
We chose the gray-box attack model presented in the paper QUID attack by Kundu and Ghosh (2024)\cite{kundu2024adversarialpoisoningattackquantum} , because it aligns well with the practical constraints of cloud-based quantum computing environments. In these settings, adversaries realistically have access to the data encoding circuits and training datasets. Still, they are unlikely to obtain full control over the training process or the complete model details. This grey-box scenario captures a realistic threat vector where an adversary within the cloud provider or intercepting communication can exploit quantum-specific vulnerabilities, such as manipulating encoded quantum states, without requiring access to the quantum hardware’s internal operations or gradients.

On the other hand, the assumptions underlying the poisoning attack in the Trusted-AI/adversarial-robustness-toolbox\cite{adversarial_robustness_toolbox} code for SVMs are less practical for real-world versions for quantum systems. The white-box model assumes complete access to the classifier, including its gradients, decision boundaries, and support vectors. Such access is rarely feasible in cloud-based quantum systems, where these details are abstracted away by the backend. Additionally, the need for iterative retraining in the SVM poisoning attack is computationally expensive and unrealistic for quantum systems, which are inherently resource-constrained. Therefore, adopting a grey-box model as in the QUID framework better reflects practical adversarial capabilities and aligns with the robustness challenges specific to the quantum domain.

The code of this experiment is located in the notebook\\
\href{https://github.com/Sultanow/quantum/blob/main/qsvm_svm_poisoning/QSVM_SVM_Poisioning.ipynb}{QSVM\_SVM\_Poisioning.ipynb} on GitHub.

\section{New QUID-Based Label Poisoning: SVM vs. QSVM}
Inspired by the Algorithm~\ref{alg:poision_qnn} presented by Kundu and Ghosh (2024) \cite{kundu2024adversarialpoisoningattackquantum}, which we present in Appendix~\ref{appx:Poisoning} (Algorithm~\ref{alg:poision_reqsvm}), we introduce a novel class of QUID-based poisoning attacks(Algorithm~\ref{alg:poision_svm} and~\ref{alg:poision_qsvm}). These attacks target both Support Vector Machines (SVMs) and Quantum Support Vector Machines (QSVMs). Additionally, in anticipation of conceptual requirements, we propose a recursive version of these attacks and prove the equivalence between the recursive and standard versions (see Appendix~\ref{appx:Poisoning}-Algorithm ~\ref{alg:poision_reqsvm} as the recursive version of Algorithm~\ref{alg:poision_svm} and Algorithm ~\ref{alg:poision_reqqsvm} as the recursive version of Algorithm~\ref{alg:poision_qsvm}).

\begin{algorithm}[H]
\caption{QUID-style Label Poisoning for Classical SVM}
\label{alg:poision_svm}
\textbf{Require:} Training data $\mathcal{D}_{\text{tr}} = \{(x_i, y_i)\}_{i=1}^n$, Poison ratio $\epsilon$, Kernel function $k(x, x')$, Distance metric $d(\cdot, \cdot)$.\
\textbf{Ensure:} Poisoned dataset with modified labels.
\BlankLine
Split $\mathcal{D}_{\text{tr}}$ into clean set $\mathcal{D}_c$ and poison set $\mathcal{D}_p$ with ratio $\epsilon$\;
$\mathcal{C} \gets \text{unique}(\{y_i \mid (x_i, y_i) \in \mathcal{D}_{\text{tr}}\})$ \tcp{Set of unique classes.}
$\Phi_c \gets \{k(x, x') \mid (x, y) \in \mathcal{D}_c\}$ \tcp{Kernel-induced clean feature space.}
$\Phi_p \gets \{k(x, x') \mid (x, y) \in \mathcal{D}_p\}$ \tcp{Kernel-induced poison feature space.}
\For{$\phi_i \in \Phi_p$}{
    $D_{\text{cls}} \gets \{\}$ \tcp{Initialize dictionary for class-wise distances.}
    \For{$c \in \mathcal{C}$}{
        $\Phi_c^{(c)} \gets \{\phi \mid \phi \in \Phi_c, y = c\}$ \tcp{Features of class $c$.}
        $D_{\text{cls}}[c] \gets \frac{1}{|\Phi_c^{(c)}|} \sum_{\phi \in \Phi_c^{(c)}} d(\phi_i, \phi)$\;
    }
    $y_i^{\text{new}} \gets \arg\max_{c \in \mathcal{C}} D_{\text{cls}}[c]$ \tcp{Assign class with maximum distance.}
}
\Return{$\mathcal{D}_c \cup \{(x_i, y_i^{\text{new}}) \mid (x_i, y_i) \in \mathcal{D}_p\}$}\;
\end{algorithm}

\medskip
\begin{algorithm}[H]
\caption{QUID-style Label Poisoning for QSVM}
\label{alg:poision_qsvm}
\textbf{Require:} Training data $\mathcal{D}_{\text{tr}} = \{(x_i, y_i)\}_{i=1}^n$, Poison ratio $\epsilon$, Encoding circuit $\phi$, Distance metric $d(\cdot, \cdot)$ for density matrices.\
\textbf{Ensure:} Poisoned dataset with modified labels.
\BlankLine
Split $\mathcal{D}_{\text{tr}}$ into clean set $\mathcal{D}_c$ and poison set $\mathcal{D}_p$ with ratio $\epsilon$\;
$\mathcal{C} \gets \text{unique}(\{y_i \mid (x_i, y_i) \in \mathcal{D}_{\text{tr}}\})$ \tcp{Set of unique classes.}
$\rho_c \gets \{\phi(x) \mid (x, y) \in \mathcal{D}_c\}$ \tcp{Encoded clean states.}
$\rho_p \gets \{\phi(x) \mid (x, y) \in \mathcal{D}_p\}$ \tcp{Encoded poison states.}
\For{$\rho_i \in \rho_p$}{
    $D_{\text{cls}} \gets \{\}$ \tcp{Initialize dictionary for class-wise distances.}
    \For{$c \in \mathcal{C}$}{
        $\rho_c^{(c)} \gets \{\rho \mid \rho \in \rho_c, y = c\}$ \tcp{States of class $c$.}
        $D_{\text{cls}}[c] \gets \frac{1}{|\rho_c^{(c)}|} \sum_{\rho \in \rho_c^{(c)}} d(\rho_i, \rho)$\;
    }
    $y_i^{\text{new}} \gets \arg\max_{c \in \mathcal{C}} D_{\text{cls}}[c]$ \tcp{Assign class with maximum distance.}
}
\Return{$\mathcal{D}_c \cup \{(x_i, y_i^{\text{new}}) \mid (x_i, y_i) \in \mathcal{D}_p\}$}\;
\end{algorithm}

\section{Results and Theoretical Insights from Experiment Design}
The results of the Monte Carlo experiments, summarized in Table~\ref{tab:results}, clearly demonstrate a significant robustness advantage of QSVM over SVM. While the SVM maintained an accuracy of 50\% under both clean and poisoned conditions, the QSVM achieved and sustained a perfect accuracy of 100\%, even when exposed to adversarial poisoning. These findings highlight the inherent resilience of quantum-based models to adversarial manipulations in the evaluated scenario.

\begin{table}[h!]
\centering
\begin{tabular}{lcc}
\hline
\textbf{Model} & \textbf{Mean Accuracy (\%)} & \textbf{Standard Deviation (\%)} \\ \hline
SVM            & 50.0                        & 0.0                             \\
SVM (Poisoned) & 49.0                        & 0.0                             \\
QSVM           & 100.0                       & 0.0                             \\
QSVM (Poisoned)& 100.0                       & 0.0                             \\ \hline
\end{tabular}
\caption{Performance of SVM and QSVM under clean and poisoned data conditions with PCA=10.}
\label{tab:results}
\end{table}

The robustness of QSVM can be attributed to its reliance on fidelity-based quantum kernels, which exploit the geometric properties of quantum states in Hilbert space. These properties make it challenging for an adversary to effectively manipulate the decision boundaries. In contrast, classical SVM relies on kernel-induced feature spaces, which are more susceptible to adversarial perturbations.

A paired t-test was conducted to assess the statistical significance of the observed difference in robustness. The results (T-statistic: $\infty$, P-value: 0.000) strongly reject the null hypothesis (\(H_0\)) that QSVM and SVM have equal robustness. This finding supports the alternative hypothesis (\(H_1\)), indicating that QSVM is significantly more robust to poisoning attacks than SVM.

\subsection{Analysis of Empirical Results}
The uniform performance of QSVM across clean and poisoned conditions suggests that quantum-enhanced models may offer inherent advantages in scenarios where data integrity cannot be guaranteed. This advantage stems from two key factors:
\begin{enumerate}
    \item \textbf{Fidelity-Based Kernels:} QSVM uses fidelity as a similarity metric, which is inherently resistant to small perturbations in quantum states. This characteristic reduces the effectiveness of poisoning attacks.
    \item \textbf{Encoding Circuit Robustness:} The use of quantum feature maps ensures that adversarial samples cannot easily align with decision boundaries in the Hilbert space.
\end{enumerate}

\subsection{PCA and Simulated Qubits}
While the results strongly favour QSVM, it is important to acknowledge certain limitations:

\begin{itemize}
    \item \textbf{Quantum System Constraints:} The experiments assume an idealized fault-tolerant quantum computing (FTQC) environment with 10 qubits. Initially, the number of features in the dataset was 701, but dimensionality reduction using PCA was applied on both SVM and QSVM, reducing it to 10 features. This reduction resulted in an accuracy of 50\% for SVM but 100\% for QSVM. To address this limitation and explore the impact of more qubits for QSVM and more features for SVM, we consider the maximum number of FTQC qubits that can be simulated using Qiskit Aer on Colab. Given the available 51 GB of RAM on Colab, the maximum number of qubits (\(n\)) that can be simulated using Qiskit Aer's state vector method is determined by the memory requirement:
\[
\text{Memory (bytes)} = 16 \cdot 2^n,
\]
where \(16\) bytes are required per amplitude (8 bytes for the real part and 8 bytes for the imaginary part).

With \( 51 \, \text{GB} = 51 \times 10^9 \, \text{bytes} \), we solve:
\[
2^n \leq \frac{51 \times 10^9}{16} \approx 3.19 \times 10^9, \quad n \approx \log_2(3.19 \times 10^9) \approx 31.9.
\]

Thus, in theory, Colab can simulate a maximum of 31 qubits, as simulating 32 qubits would exceed the memory limit.
\end{itemize}
We applied this change on SVM, and the following shows the result; as it does not improve the 50\% accuracy, we did not redo the QSVM with a higher number of simulated qubits:
\begin{table}[h!]
\centering
\begin{tabular}{lcc}
\hline
\textbf{Model} & \textbf{Mean Accuracy (\%)} & \textbf{Standard Deviation (\%)} \\ \hline
SVM            & 50.0                        & 0.0                             \\
SVM (Poisoned) & 47.0                        & 0.0                             \\ \hline
\end{tabular}
\caption{Performance of SVM and QSVM under clean and poisoned data conditions with PCA=31.}
\label{tab:results2}
\end{table} 
\newpage
\section{Theoretical Framework for QSVM Resiliency Against QUID-Style Attacks}

\subsection{QSVM is resilient to poisoning }

\begin{theorem}[Single-Qubit QML Resilience]
\label{thm:single-qubit-resilience}
Consider a single-qubit quantum model described by the state
\[
  \ket{\psi} \;=\; U(\varphi,\theta,\psi)\,\ket{\psi_0},
\]
where \(U(\varphi,\theta,\psi)\in SU(2)\) is a unitary rotation given by
Equation~\eqref{eq:rotation_su2_euler} (see also 
Section~\ref{sec:qubit_rotations_su2}). Suppose an adversary injects a small
perturbation \((\epsilon_x,\epsilon_y,\epsilon_z)\) into the Euler angles.
Then, under repeated applications of \(U\), the angular discrepancies
\(\Delta_{az}(t)\) and \(\Delta_{el}(t)\) (defined in
Equation~\eqref{eq:error_propagation} and studied via
Equations~\eqref{eq:azimuthal_elevation_difference} and
\eqref{eq:euler_matrix_power}) remain strictly \emph{periodic and bounded}
over time. Consequently, the adversarial error cannot grow unboundedly,
indicating that any single-qubit QML model is resilient to unbounded
poisoning attacks.
\end{theorem}

\begin{proof}[Proof (Outline)]
\noindent
1. \emph{Bloch-Sphere Representation.}
From Equation~\eqref{eq:qubit_matrix} (Section~\ref{sec:qubit_rotations_su2}),
a single-qubit state can be viewed as 
\(\displaystyle M_q = \vec{q}\cdot \vec{\sigma}\),
where \(\vec{q} \in \mathbb{R}^3\) is the Bloch vector and
\(\vec{\sigma}=(\sigma_1,\sigma_2,\sigma_3)\) are the Pauli matrices.
Any operation \(U(\varphi,\theta,\psi)\in SU(2)\)
(see Equation~\eqref{eq:rotation_su2_euler}) thus preserves norms and merely
rotates \(\vec{q}\) on the Bloch sphere.

\smallskip
\noindent
2. \emph{Error Injection.}
Let \(\epsilon_x,\epsilon_y,\epsilon_z\) be small offsets to the Euler angles
\((\varphi,\theta,\psi)\).  Define the \emph{azimuthal} and
\emph{elevation} discrepancies \(\Delta_{az}(t)\) and \(\Delta_{el}(t)\)
after \(t\) repeated rotations, following
Equations~\eqref{eq:error_propagation}--\eqref{eq:azimuthal_elevation_difference}.
Figure~\ref{fig:error_propagation} illustrates these discrepancies over
200 consecutive cycles, confirming periodic error patterns even with nonzero
\(\epsilon_y\).

\smallskip
\noindent
3. \emph{Periodicity and Boundedness.}
Via the matrix-power formalism 
(Equations~\eqref{eq:euler_general_matrix_power}--\eqref{eq:euler_matrix_power}),
\(\Delta_{az}(t)\) and \(\Delta_{el}(t)\) reduce to trigonometric functions in
\(\sqrt{\theta^2 + (\phi+\psi)^2}\), yielding a definite period of
\(\nicefrac{2\pi}{\sqrt{\theta^2 + (\phi+\psi)^2}}\). Tables~\ref{tab:emptytable}--\ref{tab:emptytable_1}
demonstrate numerically that these discrepancies never exceed \(\pi\), and
they oscillate between finite maxima and minima.

\smallskip
\noindent
4. \emph{Empirical Observation.}
Listings~\ref{lst:error_propagation_su2_python}--\ref{lst:error_propagation_su2_mathematica}
show Python/Mathematica implementations confirming that the discrepancy
functions stay strictly bounded (see also
Figure~\ref{fig:error_propagation_python}). Thus, even under adversarial
perturbations to a single-qubit’s rotation angles, no unbounded error growth
is possible.

\smallskip
\noindent
\textbf{Conclusion.}
Since \(\Delta_{az}(t)\) and \(\Delta_{el}(t)\) cycle through a bounded range
rather than diverging, single-qubit models exhibit inherent resilience
against unbounded poisoning. This completes the proof.
\end{proof}

Now, we prove by induction that if a QSVM with $n$ qubits is resilient, then so is a QSVM with $n+1$ qubits. Our argument leverages:

\begin{itemize}
    \item The factorization of $(n+1)$-qubit circuits into tensor products and additional gates;
    \item A Lie-algebraic embedding $\mathfrak{u}(2^n)\hookrightarrow \mathfrak{u}(2^{n+1})$;
    \item Structural invariants (norm bounds, commutator properties) under partial-trace restrictions.
\end{itemize}

These ideas collectively establish that resilience is \emph{preserved} when we enlarge an $n$-qubit QSVM to $(n+1)$ qubits.

\subsubsection{Quantum Circuits and Lie Algebras}

\begin{definition}[Unitary Groups and Lie Algebras]
  For an $n$-qubit system, let $U(2^n)$ be the group of all $2^n\times2^n$ unitary matrices. Its Lie algebra is
  \[
    \mathfrak{u}(2^n) \;=\; 
    \bigl\{H\in\mathbb{C}^{2^n\times 2^n} \;\big|\; H^\dagger = -H\bigr\}.
  \]
  Any gate $G\in U(2^n)$ can be expressed as $G=\exp\bigl(iH\bigr)$ for some $H\in \mathfrak{u}(2^n)$.
\end{definition}

\begin{definition}[QSVM Resilience]
  A quantum Support Vector Machine (QSVM) with $n$ qubits is said to be \emph{resilient} to a specified class of poisoning attacks if, for every adversarially modified training dataset $\widetilde{\mathcal{D}}$, the performance degradation (relative to the baseline dataset $\mathcal{D}$) remains provably small or negligible.
\end{definition}

Denote by $P(n)$ the proposition:
\[
  P(n) \;:\; \text{``An $n$-qubit QSVM is resilient to the specified data-poisoning attacks.''}
\]
We proceed under the following assumptions:

\begin{itemize}
    \item \textbf{Base Case:} $P(1)$ holds. That is, a single-qubit QSVM is provably resilient (as per Theorem 1).
    \item \textbf{Inductive Step (Goal):} Show $P(n)\implies P(n+1)$ for general $n$.
\end{itemize}

\subsubsection{Inductive Step: From $n$ to $n+1$ Qubits}

\begin{theorem}[Inductive Resilience Extension]\label{thm:main}
Assume $P(n)$ holds; i.e., every $n$-qubit QSVM is resilient to our specified poisoning attacks(Algorithm 2 or its recursive version, Algorithm 5). Then \emph{any} $(n+1)$-qubit QSVM, formed by suitably adding a qubit (and associated gates) to an $n$-qubit system, also remains resilient.
\end{theorem}

\begin{proof}[Proof (Outline)]
We summarize the argument in four main steps; 

\begin{enumerate}
  \item \textbf{Circuit Construction.}
  An $(n+1)$-qubit QSVM can typically be written as
  \[
    U_{n+1} \;=\; 
      \bigl(U_n \otimes I_2\bigr)
      \;\times\;
      V,
  \]
  where $U_n\in U(2^n)$ is the (assumed-resilient) $n$-qubit portion, tensored with an identity on the extra qubit, and $V$ adds entangling gates or parameterized rotations on that new qubit.

  \item \textbf{Lie-Algebraic Embedding.}
  Under the map
  \[
    \iota: \;\mathfrak{u}(2^n)\;\longrightarrow\;\mathfrak{u}(2^{n+1}), 
    \quad
    H \;\mapsto\; H\otimes I_2,
  \]
  known invariants in $\mathfrak{u}(2^n)$ (e.g., commutators, norms) embed naturally into $\mathfrak{u}(2^{n+1})$. Hence, any “robustness” property that depends on these invariants remains intact when lifting from $n$ to $n+1$ qubits.

  \item \textbf{Contradiction Argument.}
  Suppose, for contradiction, that an $(n+1)$-qubit QSVM is \emph{not} resilient. Then, a poisoning strategy exists that severely degrades its performance. However, by partially tracing out (or otherwise fixing) the extra qubit, we obtain an effective $n$-qubit subsystem that would likewise be compromised. This contradicts $P(n)$.

  \item \textbf{Conclusion.}
  The contradiction forces us to conclude that the $(n+1)$-qubit QSVM cannot be significantly corrupted by the same class of attacks. Thus, $P(n+1)$ holds under the inductive hypothesis $P(n)$, completing the extension.
\end{enumerate}
\end{proof}

\subsection{Lie-Algebraic Embedding and Invariance}
\label{sec:lie}

In quantum computing, a typical QSVM circuit of depth $m$ can be written as 
\[
  U \;=\; 
  \prod_{j=1}^m \exp\bigl(i H_j\bigr),
  \quad
  H_j \in \mathfrak{u}(2^n).
\]
Moving from $n$ qubits to $n+1$ qubits naturally replaces $\mathfrak{u}(2^n)$ by $\mathfrak{u}(2^{n+1})$. Crucially, there is an injective map
\[
  \iota: \mathfrak{u}(2^n)\;\to\;\mathfrak{u}(2^{n+1}), 
  \quad
  H \;\mapsto\; H\otimes I_2,
\]
which preserves skew-Hermiticity and, more generally, many operator-theoretic invariants:

\begin{proposition}\label{prop:embedding}
If $H_1,H_2\in\mathfrak{u}(2^n)$, then 
\[
  \bigl[\,\iota(H_1), \;\iota(H_2)\bigr]
  \;=\;
  \bigl[\,H_1\otimes I_2,\; H_2\otimes I_2\bigr]
  \;=\;
  [\,H_1, H_2\,]\;\otimes\; I_2.
\]
Hence, commutator-based invariants (and analogous spectral metrics) remain unchanged under $\iota(\cdot)$.
\end{proposition}

\begin{proof}[Proof (Outline)]
Observe that $H_1\otimes I_2$ and $H_2\otimes I_2$ commute exactly as $H_1$ and $H_2$ do, with 
\[
  [\,H_1\otimes I_2,\;H_2\otimes I_2\,]
  \;=\;
  (H_1H_2 - H_2H_1)\;\otimes\;(I_2I_2)
  \;=\;
  [\,H_1,H_2\,]\;\otimes\; I_2.
\]
Any operator norm or spectral decomposition used to characterize robustness thus carries over from $H_1,H_2$ in $\mathfrak{u}(2^n)$ to $\iota(H_1), \iota(H_2)$ in $\mathfrak{u}(2^{n+1})$ without alteration. 
\end{proof}

\begin{remark}
For poisoning attacks that exploit modifications of training data encodings in these unitaries, the key insight is: if the $n$-qubit component $U_n$ cannot be corrupted, then no additional gates on qubit $(n+1)$ alone can override that resiliency. This extends directly from the existence of an $n$-qubit \emph{subcircuit} whose properties remain intact, consistent with $P(n)$.
\end{remark}

\subsubsection{Subsystem Arguments and Partial Trace}
\label{sec:technical-lemmas}
Finally, many quantum ML protocols (including QSVMs) measure only a subset of qubits or a specific observable at the end of the circuit. If a hypothetical attack successfully poisoned an $(n+1)$-qubit QSVM, then restricting to $n$ qubits (via partial trace over the additional qubit, or conceptually “ignoring” the new qubit’s effects) would lead to a breakdown in that sub-block. This contradicts $P(n)$ by assumption, confirming resilience at $(n+1)$ qubits.\\

\subsubsection{Further improvements}
\label{sec:future-steps}
A fruitful direction for further upgrades lies in systematically tuning the parameters of the new QUID attacks across various QML models. First, varying the dimensionality of the data via different numbers of PCA components can illuminate how feature-space dimensionality impacts adversarial vulnerability and robustness. Second, exploring diverse levels of the poisoning ratio \(\epsilon\) will clarify how heavily an adversary must perturb the training dataset to significantly degrade model performance. Third, adjusting kernel parameters (e.g., fidelity-based or polynomial kernels) will allow benchmarking of which kernel properties render QML systems more or less susceptible to attack. Beyond the standard QSVM, incorporating attacks against Quantum Neural Networks, PegasusQSVM, Quantum Deep Learning frameworks, and Variational Quantum Classifiers (VQC) would provide a broad comparison of the resilience of different quantum architectures. Finally, contributing these automated routines and analyses to the open-source \cite{adversarial_robustness_toolbox} would both enrich the community’s repertoire of quantum-specific adversarial methods and foster transparent, collaborative development of robust defences and evaluations in QML.

\section{Conclusion and Outlook}
\label{sec:conclusion}

Error injection is a growing concern among practitioners and theorists in classical machine learning as models become ever more critical in organizational decision-making pipelines and as data topologies become increasingly complex. Classical models often exhibit a linear propagation of errors, which can be exploited in data poisoning attacks. By contrast, in our investigations, quantum systems---particularly those governed by the $SU(2)$ framework and leveraging fidelity-based kernels---demonstrate notable resilience to error propagation. Our theoretical and empirical findings support the notion that errors in quantum machine learning models, when viewed through Euler angle discrepancies, do not grow unboundedly and, indeed, exhibit a tendency to rise and then fall periodically.

In this paper, we introduce novel QUID-style poisoning attacks for classical SVMs and QSVMs. Our numerical experiments on synthetic radar cross-section data show that while classical SVMs degrade, the QSVM retains perfect accuracy under the same adversarial conditions. We hypothesize that this robustness is intimately tied to the intrinsic properties of quantum kernels.

Furthermore, the Lie-algebraic perspective and inductive resilience arguments indicate that if single-qubit systems display bounded error propagation, then larger quantum systems (with $n$ qubits) will also maintain resilience. This inductive extension is rooted in the structure of $\mathfrak{u}(2^n)$ embeddings, partial-trace restrictions, and the compositional nature of quantum circuits. These observations lay the groundwork for a broader theoretical framework explaining why quantum machine learning architectures can be more robust to poisoning attacks than their classical counterparts.

\subsection*{Future Directions}
While our results show promise, several important directions remain to be investigated:

\begin{enumerate}
    \item \textbf{Scaling to Larger Feature Spaces and More Qubits.} Our work examined a single synthetic use case and employed PCA to keep the feature dimension manageable for both classical SVM and QSVM. Exploring larger-scale datasets and simulating up to the maximum feasible qubit limit could validate whether the observed robustness generalizes.
    \item \textbf{Adversarial Parameter Sweeps.} Varying the poisoning ratio $\epsilon$, the fidelity-based quantum kernel properties, and different classical kernels could yield deeper insights into when quantum advantages hold and under what conditions they might erode.
    \item \textbf{Extending Attacks Beyond QSVM.} Other quantum classifiers—such as Quantum Neural Networks, Variational Quantum Classifiers, and more advanced hybrid QML frameworks—could exhibit different adversarial vulnerabilities. A broad comparative study would clarify which quantum architectures confer the strongest defences by design.
    \item \textbf{Integration with Quantum Error Correction.} It remains an open question how classical and quantum error-correction protocols, when layered on top of these quantum models, might further mitigate adversarial effects.
    \item \textbf{Open-Source Tooling.} Incorporating quantum-specific adversarial routines into established libraries (e.g., the \emph{Adversarial Robustness Toolbox}) could facilitate transparent benchmarks and standardization, advancing research on quantum-safe machine learning.
\end{enumerate}

In conclusion, this work underscores that leveraging quantum mechanical properties—specifically the use of unitary transformations and fidelity-based kernels—can curb error propagation in ways distinct from classical models. Whether one views this from a geometry-of-angles standpoint or from a Lie-algebraic analysis, the cyclic growth and subsequent decay of errors appear deeply woven into the quantum computational fabric. While our experiments represent only a first step in translating these error dynamics to real-world quantum machine learning applications, the findings point to an optimistic future where QML could serve as a more robust alternative in adversarial settings. The next wave of research—empirical and theoretical—will undoubtedly refine and extend these insights, fostering a deeper understanding of quantum resilience in adversarial machine learning. 

\newpage
\appendix
\section{The algebra behind single qubit rotations}
\label{appx:single-qubit-rotations}
The study of quantum error propagation is a fundamental aspect of quantum computing, particularly in the context of noisy intermediate-scale quantum (NISQ) devices. While the practical scenarios often involve complex multi-qubit interactions and noise propagation through entangled gates, starting with single-qubit rotations serves as a mathematically grounded and accessible first step.

While some may critique this approach as overly simplistic—especially when applied to advanced domains like quantum machine learning, where noise behaves differently due to entanglement and multi-qubit interactions -- it is important to comprehend the basic mathematical principles.

By focusing initially on single-qubit errors, we isolate key mathematical properties and develop tools that are extensible to more complex scenarios. This abstraction serves as a stepping stone, offering insights into the fundamental behaviour of quantum noise and laying the groundwork for analyzing more intricate cases, such as entanglement-based error propagation in gates like CNOT.

Our work acknowledges the limitations of single-qubit models in representing real-world quantum systems. However, the following described methods provide an instructive entry point, fostering a deeper algebraic understanding that is essential for tackling the challenges posed by multi-qubit entangled systems. We bridge the gap between foundational error analysis and the broader, more complex domain of quantum error propagation in practical quantum algorithms, including QML.

\subsection{\texorpdfstring{Qubit rotations using \boldmath{$SU(2)$}}{Qubit rotations using SU(2)}}
\label{sec:qubit_rotations_su2}
Rotations of a qubit on the Bloch sphere \cite{wie2014bloch} can be described by the group $SU(2)$ whose elements are special unitary complex $2\times2$ matrices. The rotation is given by equation~\eqref{eq:rotation_su2_euler}, see \cite[p.~67]{Normand1980}:
\begin{equation}
\label{eq:rotation_su2_euler}
U(\varphi,\theta,\psi)=e^{-i\frac{\varphi}{2}\sigma_3}e^{-i\frac{\theta}{2}\sigma_2}e^{-i\frac{\psi}{2}\sigma_3}=\begin{pmatrix}
e^{-i\frac{\varphi+\psi}{2}}\cos\frac{\theta}{2} & -e^{-i\frac{\varphi-\psi}{2}}\sin\frac{\theta}{2}\\
e^{i\frac{\varphi-\psi}{2}}\sin\frac{\theta}{2} & e^{i\frac{\varphi+\psi}{2}}\cos\frac{\theta}{2}\end{pmatrix}
\end{equation}

\par\medskip
Here $0\le\varphi\le2\pi$, $0\le\theta\le\pi$, $0\le\psi\le4\pi$ are the Euler angles and $\sigma_1=\big(\begin{smallmatrix}0 & 1\\1 & 0\end{smallmatrix}\big)$, $\sigma_2=\big(\begin{smallmatrix}0 & -i\\i & 0\end{smallmatrix}\big)$, $\sigma_3=\big(\begin{smallmatrix}1 & 0\\0 & -1\end{smallmatrix}\big)$ are the standard Pauli matrices.

A qubit (on the Bloch sphere) can be expressed as a Cartesian vector
\begin{equation}
\vec{q} = (\sin\theta_{el}\cos\varphi_{az},\sin\theta_{el}\sin\varphi_{az},\cos\theta_{el})
\end{equation} where $\varphi_{az}$ is the azimuthal angle and $\theta_{el}$ is the elevation angle of the vector $\vec{q}$ \cite[p.~3]{Yepez2013}. Moreover, using the Pauli spin vector $\vec{\sigma}=\left(\sigma_1,\sigma_2,\sigma_3\right)$, a qubit can also be expressed in matrix form given by equation~\eqref{eq:qubit_matrix}, see \cite[p.~3]{Yepez2013}.

\begin{flalign}
\label{eq:qubit_matrix}
M_q=\vec{q}\cdot\vec{\sigma}&=\sigma_1\sin\theta_{el}\cos\varphi_{az}+\sigma_2\sin\theta_{el}\sin\varphi_{az}+\sigma_3\cos\theta_{el}\\&=\begin{pmatrix}
\cos\theta_{el} & e^{-i\varphi_{az}}\sin{\theta_{el}}\\\notag
e^{i\varphi_{az}}\sin\theta_{el} & -\cos\theta_{el}\end{pmatrix}
\end{flalign}

Using this representation, the rotation of a qubit by an angle $\theta$ about an arbitrary axis $\hat n$ with $\left|\hat n\right|=1$ is then given by 
\begin{align}
M_{q'} & = U_{\hat n}(\theta)\cdot M_q\cdot U_{\hat n}^{\dagger}(\theta) \label{eq:rotation_su2_axis}
\intertext{with}
U_{\hat n}(\theta) & = e^{-i\frac{\theta}{2}\hat n\cdot\vec{\sigma}}=\sigma_0\cos\frac{\theta}{2}-i(\hat n\cdot\vec{\sigma})\sin\frac{\theta}{2}
\end{align}
where $\sigma_0=\big(\begin{smallmatrix}1 & 0\\0 & 1\end{smallmatrix}\big)$ denotes the $2\times2$ identity matrix and $(\hat n\cdot\vec{\sigma})^2=\sigma_0$, see \cite[p.~3]{Yepez2013}.


These fundamental formulas are already sufficient to implement $SU(2)$ rotations of the qubits on a Bloch sphere. Code for a Python-based implementation is provided in Appendix~\ref{appx:impl-rotations-su2} and discussed with reference to the above-given equations.

\subsection{Qubit rotations using the Euler Matrix}
\label{sec:qubit_rotations_euler_matrix}

Rotating a qubit in its Cartesian representation $\vec{q}\in\mathbb{R}^3$ by the Euler angles $\varphi$, $\theta$, and $\psi$ can be expressed as multiplication with the \href{https://reference.wolfram.com/language/ref/EulerMatrix.html}{Euler matrix} $S(\varphi,\theta,\psi)$, see \cite[p. 141]{Arfken2013}. That is

\begin{flalign}
\label{eq:rotation_euler_matrix}
q'&=q\cdot S(\varphi,\theta,\psi)\\\notag
S(\varphi,\theta,\psi)&={\textstyle\begin{pmatrix}
\cos\psi\cos\theta\cos\varphi-\sin\psi\sin\varphi & \cos\psi\cos\theta\sin\varphi+\sin\psi\cos\varphi & -\cos\psi\sin\theta\\
-\sin\psi\cos\theta\cos\varphi-\cos\psi\sin\varphi & -\sin\psi\cos\theta\sin\varphi+\cos\psi\cos\varphi & \sin\psi\sin\theta\\
\sin\theta\cos\varphi & \sin\theta\sin\varphi & \cos\theta
\end{pmatrix}}\\\notag
&=\textstyle\begin{pmatrix}
\cos\psi & \sin\psi & 0\\
-\sin\psi & \cos\psi & 0\\
0 & 0 & 1
\end{pmatrix}\begin{pmatrix}
\cos\theta & 0 & -\sin\theta\\
0 & 1 & 0\\
\sin\theta & 0 & \cos\theta
\end{pmatrix}\begin{pmatrix}
\cos\varphi & \sin\varphi & 0\\
-\sin\varphi & \cos\varphi & 0\\
0 & 0 & 1
\end{pmatrix}
\\\notag
&=S_3(\psi)S_2(\theta)S_1(\varphi)
\end{flalign}

Code for a Python-based implementation of this approach is provided in Appendix~\ref{appx:impl-rotations-euler-matrix}.

\section{Error Propagation in Matrix Multiplications}
\label{sec:error_propagation}

Let $f:\mathbb{R}^3\times\mathbb{R}^3\times\mathbb{R}^3\to \mathbb{R}^2$ a function that takes two vectors $\vec{v},\vec{v}_{err}\in\mathbb{R}^3$ and the triple $(\varphi,\theta,\psi)\in\mathbb{R}^3$ of Euler angles as input and produces a real pair containing the two discrepancies, namely those between the azimuthal and the elevation angles of the vectors $\vec{v},\vec{v}_{err}$ in the spherical coordinate system
\begin{equation}
\label{eq:error_propagation}
f \bigl(\vec{v},\vec{v}_{err},(\varphi,\theta,\psi) \bigr) = \bigl( \Delta_{az}, \Delta_{el} \bigr)
\end{equation}

We rotate $\vec{v}$ and $\vec{v}_{err}$ synchronously by the Euler angles $(\varphi,\theta,\psi)$ and obtain the rotated vectors $\vec{w}$ and $\vec{w}_{err}$. Given these rotated vectors, we determine the azimuth angular discrepancy $\Delta_{az}$ and the elevation angular discrepancy $\Delta_{el}$ as follows:

\begin{flalign*}
\Delta_{az}(\vec{w},\vec{w}_{err}) 
& = \min \Bigl\{ \lvert \varphi_{az}(\vec{w}_{err})-\varphi_{az}(\vec{w}) \rvert, 2\pi - \lvert \varphi_{az}(\vec{w}_{err})-\varphi_{az}(\vec{w}) \rvert \Bigr\} \\
\Delta_{el}(\vec{w},\vec{w}_{err})
& = \min \Bigl\{ \lvert \theta_{el}(\vec{w}_{err})-\theta_{el}(\vec{w}) \rvert, 2\pi - \lvert \theta_{el}(\vec{w}_{err})-\theta_{el}(\vec{w})\rvert \Bigl\}
\end{flalign*}

\begin{figure}[H]
	\centering
	\includegraphics[clip, trim=0cm 0cm 0cm 0cm, width=0.6\textwidth]{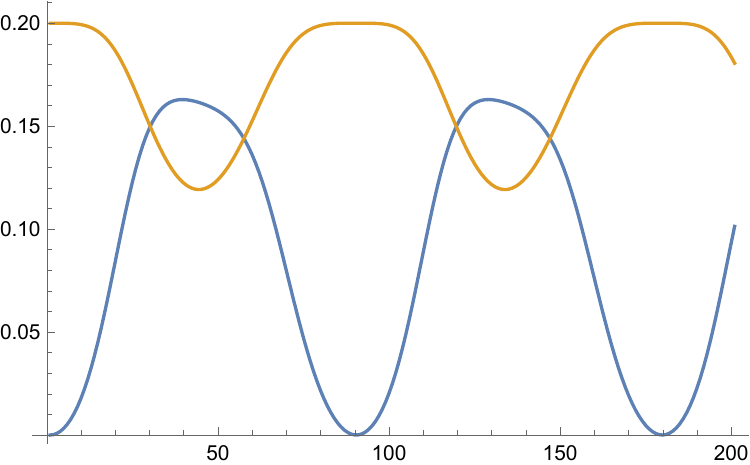}
	\caption{Curves of the error propagation (azimuth/blue and elevation/orange) due to 200 qubit rotations by $\nicefrac{\pi}{100}$.  From the physics point of view, these rotations can be considered cycles.}
	\label{fig:error_propagation}
\end{figure}

For instance, when evaluating $f$ on $\vec{v}=(1,0,0)$, $\vec{v}_{err}=(0.98,0,-0.19866933)$\footnote{The vector $\vec{v}_{err}$ results from the single rotation of vector $\vec{v}=(1,0,0)$ by the Euler angles $(\epsilon_x,\epsilon_y,\epsilon_z)=(0,0.2,0)$.} and $(\varphi,\theta,\psi)=(\nicefrac{\pi}{100},\nicefrac{\pi}{100},\nicefrac{\pi}{100})$, we obtain results as shown in Figure~\ref{fig:error_propagation}. It plots the azimuthal difference (blue) and elevation difference (orange) between both vectors $\vec{v}$ and $\vec{v}_{err}$ that are synchronously rotated $200$ times by an angle of $\nicefrac{\pi}{100}$.

There are two possibilities to generate curves as shown in Figure~\ref{fig:error_propagation}. They can first of all be obtained by applying the $SU(2)$ rotation using equation~\eqref{eq:rotation_su2_euler} that yields the rotated vectors $\vec{w}=U(\varphi,\theta,\psi)\cdot (\vec{v}\cdot\vec{\sigma})\cdot U(\varphi,\theta,\psi)^{\dagger}$ and $\vec{w}_{err}=U(\varphi,\theta,\psi)\cdot (\vec{v}_{err}\cdot\vec{\sigma})\cdot U(\varphi,\theta,\psi)^{\dagger}$, see Listing~\ref{lst:error_propagation_su2_python} (Python) and Listing~\ref{lst:error_propagation_su2_mathematica} (Mathematica).

The second, more convenient way is to use the Euler matrix \cite{quintana2018euler}. In this case we refer to equation~\eqref{eq:rotation_euler_matrix} and obtain the rotated vectors by the matrix products $\vec{w}=\vec{v}\cdot S(\varphi,\theta,\psi)$ and $\vec{w}_{err}=\vec{v}_{err}\cdot S(\varphi,\theta,\psi)$, see Listing~\ref{lst:error_propagation_euler_matrix_mathematica} (Mathematica).

The two functions that return the azimuthal and the elevation angle difference between both vectors $\vec{v}$, $\vec{v}_{err}$ that are synchronously, iteratively rotated $t$ times by a step angle of $\nicefrac{2\pi}{s}$ are:

\begin{flalign}
\label{eq:azimuthal_elevation_difference}
\Delta_{az}(\vec{v},\vec{v}_{err},t,s)&=\Delta_{az}\left(\vec{v}\cdot S\left(\nicefrac{2\pi}{s},\nicefrac{2\pi}{s},\nicefrac{2\pi}{s}\right)^t,\vec{v}_{err}\cdot S\left(\nicefrac{2\pi}{s},\nicefrac{2\pi}{s},\nicefrac{2\pi}{s}\right)^t\right)\\\notag
\Delta_{el}(\vec{v},\vec{v}_{err},t,s)&=\Delta_{el}\left(\vec{v}\cdot S\left(\nicefrac{2\pi}{s},\nicefrac{2\pi}{s},\nicefrac{2\pi}{s}\right)^t,\vec{v}_{err}\cdot S\left(\nicefrac{2\pi}{s},\nicefrac{2\pi}{s},\nicefrac{2\pi}{s}\right)^t\right)
\end{flalign}

The blue curve in Figure~\ref{fig:error_propagation} is given by $\Delta_{az}(\vec{v},\vec{v}_{err},t,200)$ and the orange curve by $\Delta_{el}(\vec{v},\vec{v}_{err},t,200)$ where $t$ is iterated from $0$ to $200$. That is we rotate the vectors $\vec{v}$, $\vec{v}_{err}$ by $\varphi=\theta=\psi=\nicefrac{\pi}{100}$, the new vectors resulting from this rotation we rotate again by $\nicefrac{\pi}{100}$ and so forth -- and this a total of $200$ times (see Listing~\ref{lst:error_propagation_euler_matrix_mathematica}).

For simplicity and to facilitate periodic behaviour analysis, we use equal values for all three Euler angles in our initial investigations. This choice allows us to focus on the overall pattern of error propagation without introducing additional complexity from different rotation rates around different axes. It provides a clear baseline for understanding error behaviour. Later in Section \ref{sec:error_propagation_analysis}, we explore cases with different Euler angles to find maximal error conditions.

For this analysis, let us by convention, set the initial vector $\vec{v}=(1,0,0)$ and the rotated (manipulated) vector $\vec{v}_{err}=\vec{v}\cdot S(\epsilon_x,\epsilon_y,\epsilon_z)$ and parameterize both functions~\eqref{eq:azimuthal_elevation_difference} with these angles and with the run variable $t$ and the step variable $s$: $\Delta_{az}(\epsilon_x,\epsilon_y,\epsilon_z,t,s)$, $\Delta_{el}(\epsilon_x,\epsilon_y,\epsilon_z,t,s)$. We can use those two functions to determine how both vectors $\vec{v}$, $\vec{v}_{err}$ differ in their azimuthal and elevation angles after $t$ rotations each by $\nicefrac{2\pi}{s}$. The blue and orange curves do not change their shape if, for example, instead $s=200$ we now set $s=100$ and let $t$ run from $0$ to $100$. The shape of both curves depends exclusively on the three angles $\epsilon_x$, $\epsilon_y$, $\epsilon_z$.

How do we determine a real function for the blue curve and for the orange curve, which takes only one variable $t$ rather than the two variables $s$ and $t$? For this, we generate a finite rotation $S_P$ around the angle $t$ by executing infinitesimal rotations successively and calculate the following limit for a general case:

\begin{equation}
\label{eq:euler_general_matrix_power}
\begin{aligned}
& \lim_{s\to\infty} S\left(\frac{\phi}{s},\frac{\theta}{s},\frac{\psi}{s}\right)^{s \cdot t} = \lim_{s\to\infty} S\left(\frac{\phi \cdot t}{s},\frac{\theta \cdot t}{s},\frac{\psi \cdot t}{s}\right)^s = S_P(t, \phi, \theta, \psi) \\[1em]
& = \scriptscriptstyle\begin{pmatrix}
    \cosh\left(t \cdot \sqrt{-\theta^2 - (\phi + \psi)^2}\right) & 
    -\frac{(\phi + \psi) \cdot \sinh\left(t \cdot \sqrt{-\theta^2 - (\phi + \psi)^2}\right)}{\sqrt{-\theta^2 - (\phi + \psi)^2}} & 
    \frac{\theta \cdot \sinh\left(t \cdot \sqrt{-\theta^2 - (\phi + \psi)^2}\right)}{\sqrt{-\theta^2 - (\phi + \psi)^2}} \\
    \frac{(\phi + \psi) \cdot \sinh\left(t \cdot \sqrt{-\theta^2 - (\phi + \psi)^2}\right)}{\sqrt{-\theta^2 - (\phi + \psi)^2}} & 
    \frac{\theta^2 + (\phi + \psi)^2 \cdot \cosh\left(t \cdot \sqrt{-\theta^2 - (\phi + \psi)^2}\right)}{\theta^2 + (\phi + \psi)^2} & 
    -\frac{\theta \cdot (\phi + \psi) \cdot \left(-1 + \cosh\left(t \cdot \sqrt{-\theta^2 - (\phi + \psi)^2}\right)\right)}{\theta^2 + (\phi + \psi)^2} \\[10pt]
    -\frac{\theta \cdot \sinh\left(t \cdot \sqrt{-\theta^2 - (\phi + \psi)^2}\right)}{\sqrt{-\theta^2 - (\phi + \psi)^2}} & 
    -\frac{\theta \cdot (\phi + \psi) \cdot \left(-1 + \cosh\left(t \cdot \sqrt{-\theta^2 - (\phi + \psi)^2}\right)\right)}{\theta^2 + (\phi + \psi)^2} & 
    \frac{(\phi + \psi)^2 + \theta^2 \cdot \cosh\left(t \cdot \sqrt{-\theta^2 - (\phi + \psi)^2}\right)}{\theta^2 + (\phi + \psi)^2}
\end{pmatrix}
\end{aligned}
\end{equation}

and for our case, the limit tends to be:

\begin{flalign}
\label{eq:euler_matrix_power}
\lim_{s\to\infty} S\left(\frac{1}{s},\frac{1}{s},\frac{1}{s}\right)^{s\cdot t}&=\lim_{s\to\infty} S\left(\frac{t}{s},\frac{t}{s},\frac{t}{s}\right)^s=S_P(t)\\\notag
&=\left(
\begin{array}{ccc}
\cos\left(\sqrt{5}t\right)&-\frac{2\sin\left(\sqrt{5}t\right)}{\sqrt{5}}&\frac{\sin\left(\sqrt{5}t\right)}{\sqrt{5}}\\
\frac{2\sin\left(\sqrt{5}t\right)}{\sqrt{5}} & \frac{1}{5}\left(4\cos\left(\sqrt{5}t\right)+1\right) & -\frac{2}{5}\left(\cos\left(\sqrt{5}t\right)-1\right)\\
-\frac{\sin\left(\sqrt{5}t\right)}{\sqrt{5}} & -\frac{2}{5}\left(\cos\left(\sqrt{5}t\right)-1\right) & \frac{1}{5}\left(\cos\left(\sqrt{5}t\right)+4\right) \\
\end{array}
\right)
\end{flalign}

As a result, we get the following two functions, which (by setting again $\epsilon_x=0$, $\epsilon_y=0.2$, $\epsilon_z=0$) generate exactly the same blue and orange curve plotted with Figure~\ref{fig:error_propagation} (Listing~\ref{lst:error_propagation_limit_euler_matrix_power_mathematica}):

\begin{flalign}
\label{eq:azimuthal_elevation_difference_limit_euler_matrix_power}
\Delta_{az}(\epsilon_x,\epsilon_y,\epsilon_z,t)&=\Delta_{az}\left(\Big(\begin{smallmatrix}1\\0\\0\end{smallmatrix}\Big)\cdot S_P(t),\Big(\begin{smallmatrix}1\\0\\0\end{smallmatrix}\Big)\cdot S(\epsilon_x,\epsilon_y,\epsilon_z)\cdot S_P(t)\right)\\\notag
&=\Delta_{az}\left(\vec{v}\cdot S_P(t),\vec{v}_{err}\cdot S_P(t)\right)\\\notag
\Delta_{el}(\epsilon_x,\epsilon_y,\epsilon_z,t)&=\Delta_{el}\left(\Big(\begin{smallmatrix}1\\0\\0\end{smallmatrix}\Big)\cdot S_P(t),\Big(\begin{smallmatrix}1\\0\\0\end{smallmatrix}\Big)\cdot S(\epsilon_x,\epsilon_y,\epsilon_z)\cdot S_P(t)\right)\\\notag
&=\Delta_{el}\left(\vec{v}\cdot S_P(t),\vec{v}_{err}\cdot S_P(t)\right)\\\notag
\end{flalign}

Let us consider rotations around infinitesimally small angles $\partial{t}$ and represent these infinitesimal rotations as $S_P(\partial{t})=I+\partial{t}~J$, where $I$ is the identity matrix and $J$ the generator of the infinitesimal rotation:

\[
J=\left.\frac{\partial{S_P(t)}}{\partial{t}}\right|_{t=0}=\lim_{t\to0}\frac{S_P(t)-I}{t}=S_P^{-1}\frac{\partial{S_P}}{\partial{t}}
\]

In reverse, we can express the finite rotation $S_P$ around the angle $t$ as follows:

\[
S_P(t)=\lim_{s\to\infty}S_P\left(\frac{t}{s}\right)^s=\lim_{s\to\infty}\left(I+\frac{t}{s}J\right)^s=\exp{(tJ)}
\]

In a general case, the generator turns out to be a traceless matrix \cite{chruscinski2021constraints}:

\[
J=\begin{pmatrix}
0 & -\phi-\psi & \theta\\
\phi+\psi & 0 & 0\\
-\theta & 0 & 0
\end{pmatrix}
\]

Along with that, the time period associated with the elevation error and azimuthal error can be expressed as:

\[
T = \frac{2\pi}{\sqrt{\theta^2 + (\phi + \psi)^2}}
\] 

In our case the generator is:

\[
J=\begin{pmatrix}
0 & -2 & 1\\
2 & 0 & 0\\
-1 & 0 & 0
\end{pmatrix}
\]

The eigenvalues of $J$ in the general case tends to be $0$ and $\pm i\sqrt{\theta^2+(\psi+\phi)^2}$ and for our case they are $0$ and $\pm i\sqrt{5}$. Let us consider the quadruple $(M,G,E,\Phi)$, where $M$ is the phase space containing three-dimensional rotation matrices, $G$ the group of real numbers (as a model for the progression of time), $E$ is the subset $E\subseteq G\times M$, and $\Phi:E\rightarrow M$ is an operation of the group $G$ on $M$ with $\Phi(0,x)=x$ for all $x\in M$ and $\Phi(s,\Phi(t,x))=\Phi(s\cdot t,x)$ for all $x\in M$ and for all $s,t\in G$. Then $(M,G,E,\Phi)$ is a dynamical system and $\Phi$ is the flow on $M$, see \cite[p.~131-140]{Wirsching2004}.




\section{Error propagation analysis}
\label{sec:error_propagation_analysis}

In the following, we will describe the behaviour of the curves given by $\Delta_{az}$ and $\Delta_{el}$, including the periodicity and the minimum and maximum difference between the angles of the two simultaneously rotating vectors.

\subsection{Periodicity}
The periodicity of both curves for the above case, namely the curves given by $\Delta_{az}$ and $\Delta_{el}$, is always $\nicefrac{2\pi}{\sqrt{5}}$, no matter which angles $\epsilon_x$, $\epsilon_y$, $\epsilon_z$ we choose. We obtained this value by

\code{FunctionPeriod[elevationErrorSimplified[x, y, z, t], t]} and

\code{FunctionPeriod[azimuthalErrorSimplified[x, y, z, t], t]}. 

This can be proved by the equalities $\Delta_{az}(\epsilon_x,\epsilon_y,\epsilon_z,t)=\Delta_{az}(\epsilon_x,\epsilon_y,\epsilon_z,t+\nicefrac{2\pi}{\sqrt{5}})$ and $\Delta_{el}(\epsilon_x,\epsilon_y,\epsilon_z,t)=\Delta_{el}(\epsilon_x,\epsilon_y,\epsilon_z,t+\nicefrac{2\pi}{\sqrt{5}})$, referring to the equations~\eqref{eq:azimuthal_elevation_difference_limit_euler_matrix_power}.

\subsection{Maximum and minimum angular difference}
To obtain the maximum and minimum that an elevation or azimuthal angle between both rotating vectors $\vec{v}, \vec{v_{err}}$ can have, we perform a numerical search for the maximum and minimum values of the functions~\eqref{eq:azimuthal_elevation_difference_limit_euler_matrix_power}:

\begin{listing}[H]
\begin{minted}[bgcolor=bg, frame=leftline, linenos, texcl=false, framesep=2mm, breaklines, tabsize=2]{mathematica}
elevationErrorSimplified[errx_, erry_, errz_, t_] := 
  angularErrorSimplified[{1, 0, 0}, {1, 0, 0} . 
    EulerMatrix[{errx, erry, errz}], t, 2];
    
azimuthalErrorSimplified[errx_, erry_, errz_, t_] := 
  angularErrorSimplified[{1, 0, 0}, {1, 0, 0} . 
    EulerMatrix[{errx, erry, errz}], t, 3];
    
    
NMaximize[{elevationErrorSimplified[x, y, z, t], 0 < x < 2 Pi && 0 < y < 2 Pi && 0 < z < 2 Pi && 0 < t < 2 Pi}, {x, y, z, t}]
NMaximize[{azimuthalErrorSimplified[x, y, z, t], 0 < x < 2 Pi && 0 < y < 2 Pi && 0 < z < 2 Pi && 0 < t < 2 Pi}, {x, y, z, t}]
\end{minted}
\caption{Calculate the maximum elevation and azimuthal difference between both rotating vectors $\vec{v}$ and $\vec{v}_{err}$ (based on Listing~\ref{lst:error_propagation_limit_euler_matrix_power_mathematica})}
\label{lst:findmaximum_mathematica}
\end{listing}

The maximum elevation difference between $\vec{v}$ and $\vec{v}_{err}$ is $2.0344424161175363$ which occur for $\epsilon_x=5.77302981892348$, $\epsilon_y=4.3028173057175945$, $\epsilon_z=4.602388461805818$, $t=3.511308418550631$.

Analogously we obtain the maximum azimuthal difference between both vectors which is $\pi$ and occur for $\epsilon_x=1.697427669696458$, $\epsilon_y=6.202505314481809$, $\epsilon_z=1.4563517625363593$, $t=3.952403209518636$.

The minimum elevation difference is $\approx0$ which occur for $\epsilon_x=1.6952986881010703$, $\epsilon_y=1.546558501896838$, $\epsilon_z=4.608182273912689$, $t=4.061805642653761$.

The minimum azimuthal difference between $\vec{v}$ and $\vec{v}_{err}$ is $\approx0$ which occur for $\epsilon_x=5.017163536971859$, $\epsilon_y=3.9579710903948024$, $\epsilon_z=1.7852930269424405$, $t=2.911162125823624$.

\begin{table}[H]
\centering
\begin{tabular}{|c|c|c|c|c|}
\hline
\tiny{Azim./Elev. diff.}($\vec{v}$ and $\vec{v}_{err}$) & t & $\epsilon_x$ & $\epsilon_y$ & $\epsilon_z$\\
\hline
*\tiny{2.0344424161175363} & \tiny{$3.511308418550631$} &\tiny{$5.77302981892348$}, & \tiny{$4.3028173057175945$}, & \tiny{$4.602388461805818$} \\
\hline
**$\pi$ & \tiny{$3.952403209518636$} &\tiny{$1.697427669696458$}, & \tiny{ $6.202505314481809$}, & \tiny{$1.4563517625363593$}  \\
\hline
+\tiny{0} &  \tiny{$4.061805642653761$} &  \tiny{$1.6952986881010703$} & \tiny{$1.546558501896838$} & \tiny{$4.608182273912689$}   \\
\hline
++\tiny{ $\approx0$} &  \tiny{$2.911162125823624$} &  \tiny{$5.017163536971859$}  & \tiny{$3.9579710903948024$} & \tiny{$1.7852930269424405$}  \\
\hline

\end{tabular}
\caption{
    \\ \tiny{*} maximum elevation difference -  \tiny{**} maximum azimuthal difference \\
       \tiny{+} minimum elevation difference -  \tiny{++} minimum azimuthal difference  
       }
\label{tab:emptytable}
\end{table}

When we use the vector \code{\{0, 0, 1\}} instead of \code{\{1, 0, 0\}} in Listing~\ref{lst:findmaximum_mathematica}, then the maximum and minimum values change as given in the following. The maximum elevation difference is $\pi$ which occur for $\epsilon_x=3.373959984091561$, $\epsilon_y=\pi$, $\epsilon_z=2.8479974737420557$, $t=\nicefrac{2\pi}{\sqrt{5}}$.

Analogously we obtain the maximum azimuthal difference between both vectors, which is $\pi$ as well and occur for $\epsilon_x=3.117350980755296$, $\epsilon_y=\pi$, $\epsilon_z=4.292091776219165$, $t=4.555204697315162$.

The minimum elevation difference is $\approx0$ which occur for $\epsilon_x=5.651168952220518$, $\epsilon_y=6.283102504046935$, $\epsilon_z=0.8902613576650594$, $t=6.277761692229545$.

The minimum azimuthal difference between $\vec{v}$ and $\vec{v}_{err}$ is $\approx0$ as well which occur for $\epsilon_x=1.626524700226482$, $\epsilon_y=2\pi$, $\epsilon_z=0.18398822116968577$, $t=6.252198460931593$.

\begin{table}[H]
\centering
\begin{tabular}{|c|c|c|c|c|}
\hline
\tiny{Azim./Elev. diff.}($\vec{v}$ and $\vec{v}_{err}$) & t & $\epsilon_x$ & $\epsilon_y$ & $\epsilon_z$\\
\hline
\tiny{*}\tiny{$\pi$} & \tiny{$\nicefrac{2\pi}{\sqrt{5}}$} &\tiny{$3.373959984091561$}, & \tiny{$\pi$} & \tiny{$2.8479974737420557$} \\
\hline
\tiny{**}\tiny{$\pi$} & \tiny{$4.555204697315162$} &\tiny{$3.117350980755296$}, & \tiny{$\pi$} & \tiny{$4.292091776219165$} \\
\hline
\tiny{+}\tiny{ $\approx0$} &  \tiny{$6.277761692229545$} &  \tiny{$5.651168952220518$} & \tiny{$6.283102504046935$} & \tiny{$0.8902613576650594$}   \\
\hline
\tiny{++}\tiny{ $\approx0$} &  \tiny{$6.252198460931593$} &  \tiny{$1.626524700226482$}  & \tiny{$2\pi$} & \tiny{$0.18398822116968577$}  \\
\hline

\end{tabular}
\caption{
    \\ \tiny{*} maximum elevation difference -  \tiny{**} maximum azimuthal difference \\
       \tiny{+} minimum elevation difference -  \tiny{++} minimum azimuthal difference  
       }
\label{tab:emptytable_1}
\end{table}

\subsection{Time-Averaged Error Analysis}
While our previous analysis identified instantaneous maximum errors of $\pi$ in azimuthal difference, this may not fully represent the typical error experienced over time. A more comprehensive approach is to consider the time-averaged error over a complete period:
\begin{equation}
E_{avg}(\epsilon_x,\epsilon_y,\epsilon_z) = \frac{\sqrt{5}}{2\pi}\int_0^{2\pi/\sqrt{5}} \Delta(\epsilon_x,\epsilon_y,\epsilon_z,t) dt
\end{equation}
where $\delta$ can be either $\delta_{az}$ or $\delta_{el}$. Initial numerical evaluations suggest that time-averaged errors are lower than the instantaneous maxima found in Table. \ref{tab:emptytable}.

\subsection{Case-By-Case Analysis} Here, we conduct a detailed case-by-case analysis of error propagation examining various combinations of Euler angles. Each case assesses the maximum and minimum discrepancies in azimuthal and elevation angles, arising from comparisons between the vectors \(\vec{v}\) and \(\vec{v}_{err}\) after undergoing synchronous rotations. The analysis provides insights into how angular discrepancies evolve according to the nature of the rotations. Importantly, the periodicity in error propagation remains consistently present across all cases, though the period of the curves varies depending on the specific Euler angle configurations. The case where all Euler angles are equal has been discussed previously and will not be revisited here.

\subsubsection{Case 1: Rational Ratios of Euler Angles}

\begin{figure}[ht]
    \centering
    \begin{minipage}[b]{0.45\textwidth}
        \centering
        \includegraphics[width=\textwidth]{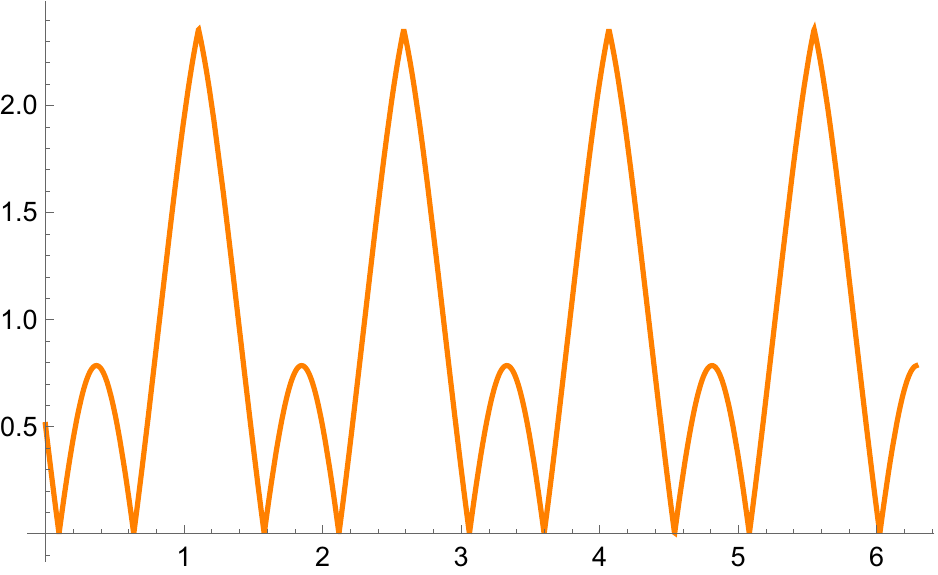}
    \end{minipage}
    \hfill
    \begin{minipage}[b]{0.45\textwidth}
        \centering
        \includegraphics[width=\textwidth]{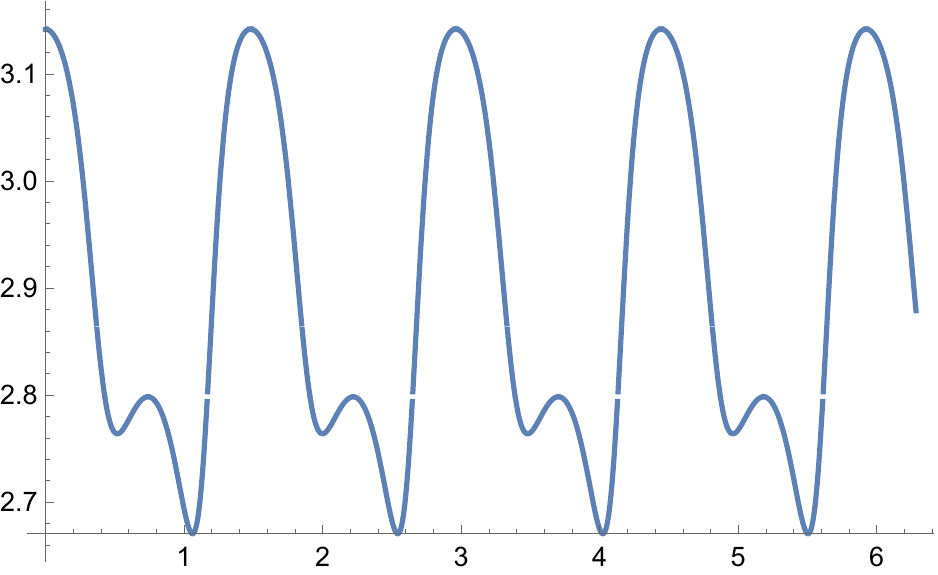} 
    \end{minipage}
    \caption{The left figure shows the Elevation Error plot, while the right shows the azimuthal error plot for Case 1, where Euler angles have a rational ratio relationship.}
    \label{rational}
\end{figure}

In this case, we explore scenarios where the Euler angles have rational ratios: specifically, a ratio of \(1:2:3\) for subcase 1 and \(3:2:1\) for subcase 2, corresponding to \(\psi\), \(\theta\), and \(\phi\). The maximum elevation error for subcase 1 is approximately \(1.76818818822050\), which is slightly lower than that of the previously discussed case. In contrast, for subcase 2, the elevation error increases to around \(2.35590586664925\), while the azimuthal errors for both subcases approach \(\pi\). The minimal errors in both cases remain negligible. The period of the curves for subcase 1 is \(\sqrt{2/13} \cdot \pi\), while for subcase 2, the period is \(\sqrt{2} \cdot \pi/3\). The periodicity of error propagation is maintained with subtle variations in the error curves, as illustrated in the figure \ref{rational}.

\subsubsection{Case 2: Two Equal Euler Angles with Rational Ratio}

\begin{figure}[ht]
    \centering
    \begin{minipage}[b]{0.45\textwidth}
        \centering
        \includegraphics[width=\textwidth]{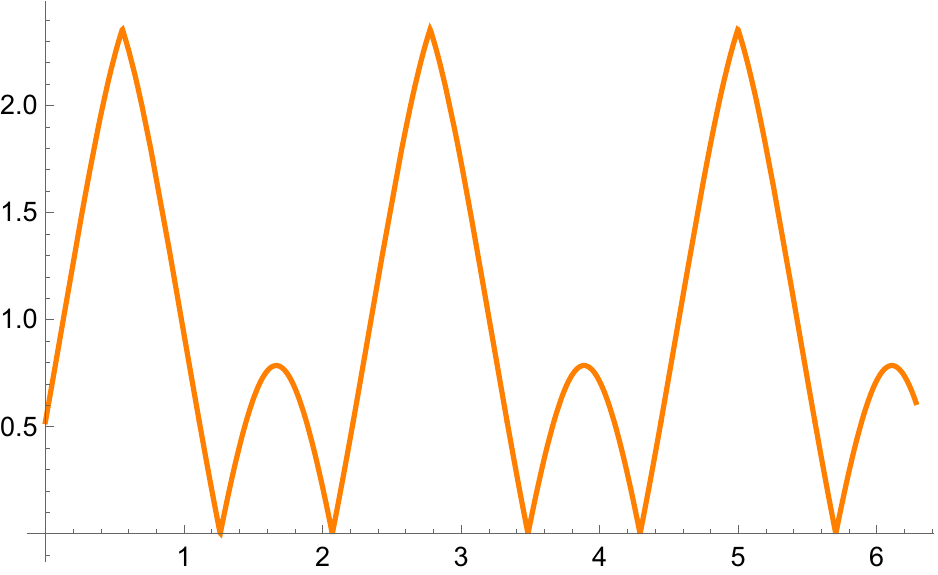}
    \end{minipage}
    \hfill
    \begin{minipage}[b]{0.45\textwidth}
        \centering
        \includegraphics[width=\textwidth]{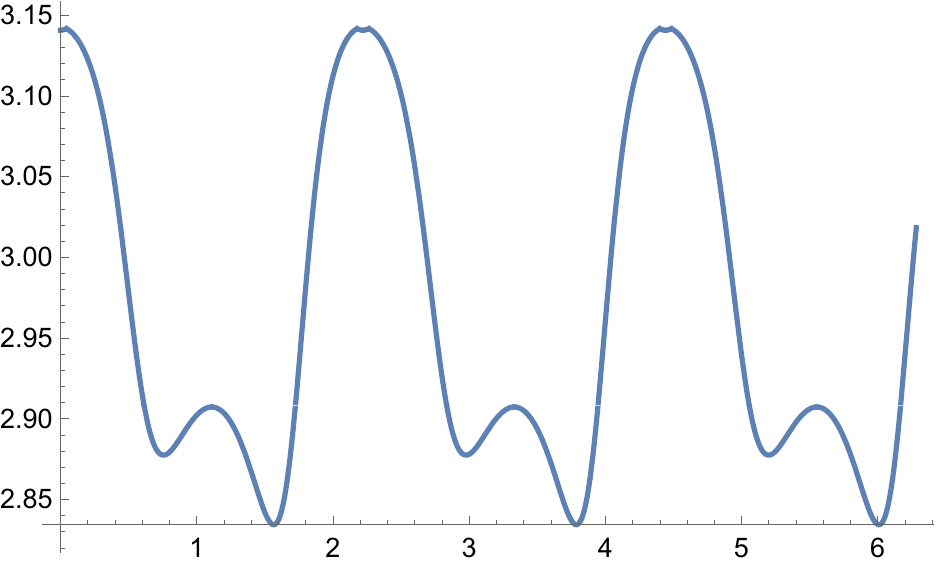} 
    \end{minipage}
    \caption{The left figure shows the Elevation Error plot, while the right shows the azimuthal error plot for Case 2, where Euler angles have a rational ratio relationship with two angles being the same.}
    \label{2rational}
\end{figure}

Here, we investigate the scenario where two Euler angles are equal, specifically in a \(1:1:2\) ratio for subcase 1 and a \(2:1:1\) ratio for subcase 2, corresponding to \(\psi\), \(\theta\), and \(\phi\). This introduces a rational ratio between the third angle in subcase 1 and the first angle in subcase 2. The maximum elevation error for subcase 1 is approximately \(1.57079632670986\), while for subcase 2, it reaches around \(2.35619448416057\). The maximum azimuthal error remains at \(\pi\), with minimal errors again approaching zero. The period of the curves for subcase 1 is \(\sqrt{2/5} \cdot \pi\), and for subcase 2, it is \(\pi/\sqrt{2}\). This case shows that, even when two angles are synchronized, the periodic behaviour of error propagation persists, with adjustments in rotation angles as shown in the figure \ref{2rational}.

\subsubsection{Case 3: Irrational Ratios of Euler Angles}

\begin{figure}[ht]
    \centering
    \begin{minipage}[b]{0.45\textwidth}
        \centering
        \includegraphics[width=\textwidth]{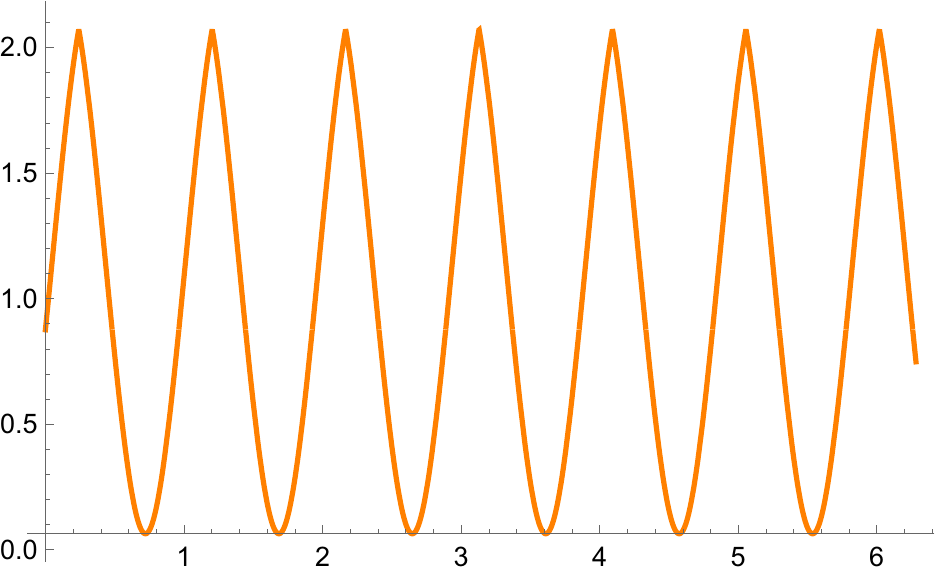}
    \end{minipage}
    \hfill
    \begin{minipage}[b]{0.45\textwidth}
        \centering
        \includegraphics[width=\textwidth]{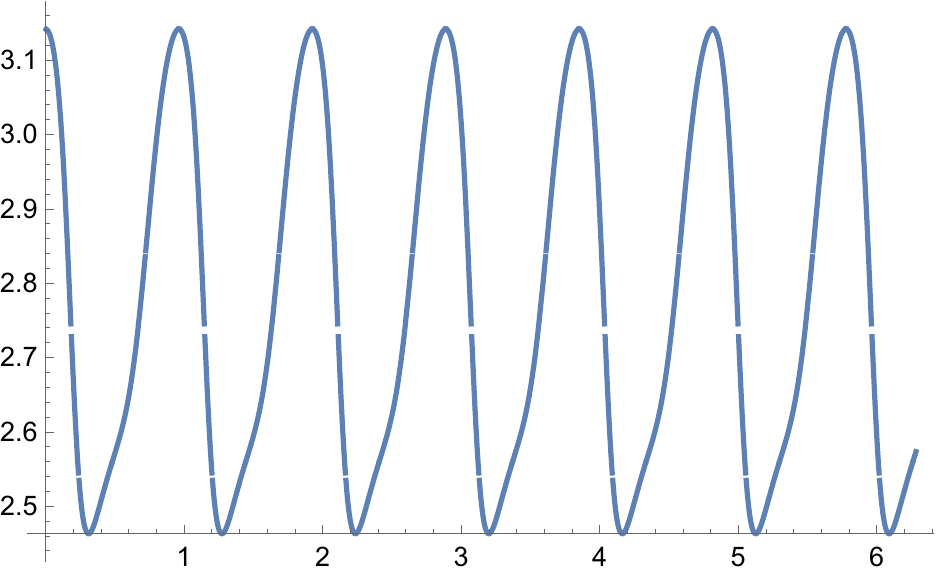} 
    \end{minipage}
    \caption{The left figure shows the Elevation Error plot, while the right shows the azimuthal error plot for Case 3, where Euler angles have an irrational ratio relationship.}
    \label{irrational}
\end{figure}

In this scenario, we select the Euler angles as \(\varphi=\pi\), \(\theta=e\), and \(\psi=3\), introducing irrational ratios among the angles. The maximum elevation error increases to approximately \(2.07317454885058\), while the maximum azimuthal error remains at \(\pi\). Although the minimal errors are small, they are not negligible. The period of the curves in this case is \(2\pi/\sqrt{\pi^2 + (e + 3)^2}\). This case illustrates the influence of irrational ratios on error propagation, resulting in more complex periodic behaviour, as depicted in the figure \ref{irrational}.

\subsubsection{Case 4: Two Equal Euler Angles with Irrational Ratio}

\begin{figure}[ht]
    \centering
    \begin{minipage}[b]{0.45\textwidth}
        \centering
        \includegraphics[width=\textwidth]{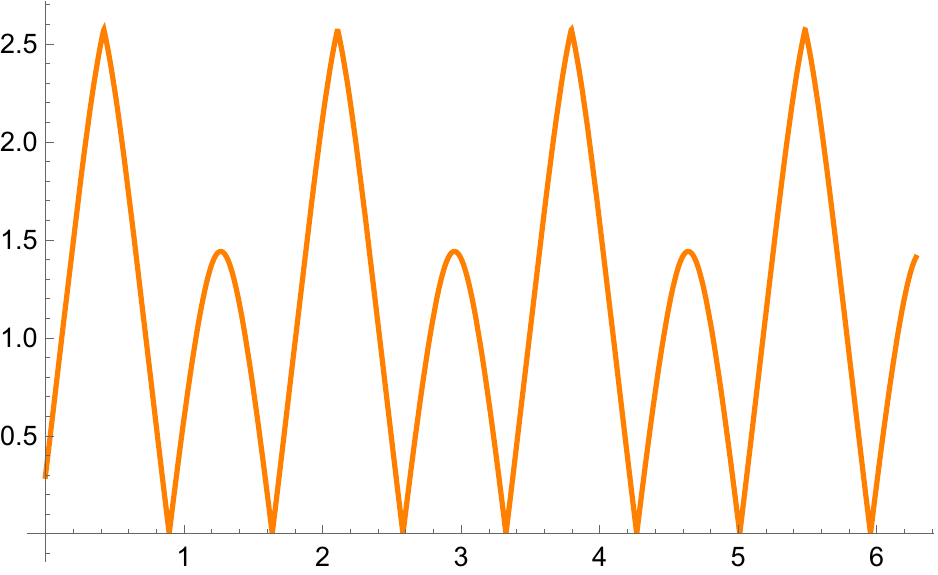}
    \end{minipage}
    \hfill
    \begin{minipage}[b]{0.45\textwidth}
        \centering
        \includegraphics[width=\textwidth]{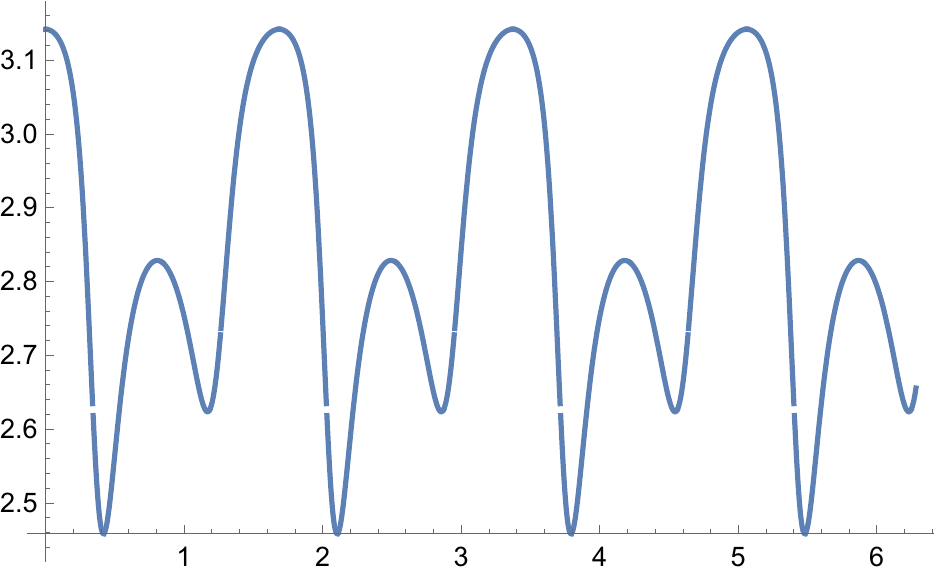} 
    \end{minipage}
    \caption{The left figure shows the Elevation Error plot, while the right shows the azimuthal error plot for Case 4, where Euler angles have an irrational ratio relationship with two angles being the same.}
    \label{2irrational}
\end{figure}

Finally, we examine a case where two Euler angles are equal, with the third angle being irrational. Specifically, we use \(1:1:\pi\) for subcase 1 and \(\pi:1:1\) for subcase 2. The maximum elevation error in subcase 1 is approximately \(1.80771464098098\), while in subcase 2, it reaches around \(2.57468030909556\). The maximum azimuthal error remains at \(\pi\). The period of the curves for subcase 1 is \(2\pi/\sqrt{1 + (\pi + 1)^2}\), while for subcase 2, it is \(2\pi/\sqrt{\pi^2 + 4}\). The minimal errors are again close to zero, with periodic patterns remaining, though more complex due to the irrational component, as illustrated in Figure \ref{2irrational}.

\section{\texorpdfstring{Implementation of $SU(2)$ rotations using Python}{Implementation of SU(2) rotations using Python}}
\label{appx:impl-rotations-su2}

In the following, we present an implementation of SU(2) rotation of qubits on a Bloch sphere based on Python. In addition to several standard libraries, especially \emph{Qiskit}, \emph{Matplotlib} and \emph{SymPy} are used (see entire notebook \href{https://github.com/Sultanow/quantum/blob/main/bloch_sphere/rotate_su2_qiskit_eldar-sultanow.ipynb}{rotate\_su2\_qiskit\_eldar-sultanow.ipynb} on GitHub).

The function \code{cartesian_to_spherical} in Listing~\ref{lst:cartesian_to_spherical} calculates the spherical coordinates $(r,\theta_{el},\varphi_{az})$ for a given cartesian vector. In line with the convention as per \emph{ISO 80000-2}, this function returns the same result as Wolfram's function \href{https://reference.wolfram.com/language/ref/ToSphericalCoordinates.html}{ToSphericalCoordinates}.

\begin{listing}[H]
\begin{minted}[bgcolor=bg, frame=leftline, linenos, texcl=false, framesep=2mm, breaklines, tabsize=2]{python}
def cartesian_to_spherical(vec):
    x = np.real(vec[0])
    y = np.real(vec[1])
    z = np.real(vec[2])
    hxy = np.hypot(x, y)
    r = np.hypot(hxy, z)
    θ = np.arctan2(hxy, z)
    φ = np.arctan2(y, x)
    return [r, θ, φ]
\end{minted}
\caption{Convert a vector from cartesian to spherical form}
\label{lst:cartesian_to_spherical}
\end{listing}

A qubit $M_q$ that is given in matrix form as per equation~\eqref{eq:qubit_matrix} can be converted to a cartesian vector $\vec{q}=(q_1,q_2,q_3)$ using the equation~\eqref{eq:qubit_matrix_to_cartesian}, see \cite[p.~79]{Normand1980}.

\begin{flalign}
\label{eq:qubit_matrix_to_cartesian}
\vec{q}\cdot\vec{\sigma}&=\frac{1}{2}\left(\left(q_1+iq_2\right)\left(\sigma_1-i\sigma_2\right)+\left(q_1-iq_2\right)\left(\sigma_1+i\sigma_2\right)+q_3\sigma_3\right)\\\notag
&=\begin{pmatrix}q_3 & q_1-iq_2\\q_1+iq_2 & -q_3\end{pmatrix}
\end{flalign}

We use this behaviour for implementing the conversion function in Listing~\ref{lst:qubit_to_cartesian}.

\begin{listing}[H]
\begin{minted}[bgcolor=bg, frame=leftline, linenos, texcl=false, framesep=2mm, breaklines, tabsize=2]{python}
def qubitmatrix_to_cartesian(M_q):
    M_q = N(M_q)
    q_1 = re((M_q[0,1] + M_q[1,0]) / 2)
    q_2 = re((M_q[1,0] - M_q[0,1]) / (2*I))
    q_3 = re(M_q[0,0])
    return np.array([q_1, q_2, q_3], dtype=np.float64)
\end{minted}
\caption{Convert a qubit given in matrix form as per equation~\eqref{eq:qubit_matrix} to cartesian form}
\label{lst:qubit_to_cartesian}
\end{listing}

\begin{listing}[H]
\begin{minted}[bgcolor=bg, frame=leftline, linenos, texcl=false, framesep=2mm, breaklines, tabsize=2]{python}
def rn_su2_euler(vec, rx, ry, rz):
    spherical_vec = cartesian_to_spherical(vec)
    θ = spherical_vec[1]
    φ = spherical_vec[2]
    
    sx = msigma(1)
    sy = msigma(2)
    sz = msigma(3)
    M_q = sin(θ)*cos(φ)*sx + sin(θ)*sin(φ)*sy + cos(θ)*sz
    U_n = Matrix([[exp(-I*(rx+rz)/2)*cos(ry/2), -exp(-I*(rx-rz)/2)*sin(ry/2)], [exp(I*(rx-rz)/2)*sin(ry/2), exp(I*(rx+rz)/2)*cos(ry/2)]])
    M_q_rotated = U_n*M_q*Dagger(U_n)
    return M_q_rotated
\end{minted}
\caption{Rotate a qubit around Euler angles as per equation~\eqref{eq:rotation_su2_euler}}
\label{lst:rotate_su2_euler}
\end{listing}

Listing~\ref{lst:album_rotate_su2_euler} applies the $SU(2)$ rotation around Euler angles that is implemented by Listing~\ref{lst:rotate_su2_euler} and plots different rotations against each other.

\begin{listing}[H]
\begin{minted}[bgcolor=bg, frame=leftline, linenos, texcl=false, framesep=2mm, breaklines, tabsize=2]{python}
fig, ax = plt.subplots(figsize = [8, 12], nrows=3, ncols=2)
fig.patch.set_facecolor('white')
[axis.set_axis_off() for axis in ax.ravel()]

rotations = [[0, 0, pi/8], [0, 0, -pi/8], [0, pi/8, 0], [0, -pi/8, 0], [0, pi/8, pi/8], [0, -pi/8, -pi/8]]
start_vec = [1, 0, 0]
num_iterations = 8
for m, rotation in enumerate(rotations):
    ax = fig.add_subplot(320+(m+1), axes_class = Axes3D)
    rot_x = rotation[0]
    rot_y = rotation[1]
    rot_z = rotation[2]
    _bloch = Bloch(axes=ax)
    _bloch.vector_color = get_gradient_colors([0, 0, 1], num_iterations)
    _bloch.vector_width = 1
    sv = []
    vec = start_vec
    sv.append(vec)
    for i in range(num_iterations):
        M_q_rotated = rn_su2_euler(vec, rot_x, rot_y, rot_z)
        vec = qubitmatrix_to_cartesian(M_q_rotated)
        sv.append(vec)

    _bloch.add_vectors(sv)
    _bloch.render()
\end{minted}
\caption{Various $SU(2)$ rotations of the vector $(1,0,0)$ around different Euler angles}
\label{lst:album_rotate_su2_euler}
\end{listing}

\begin{figure}[H]
	\centering
	\includegraphics[clip, trim=0cm 0cm 0cm 0cm, width=0.6\textwidth]{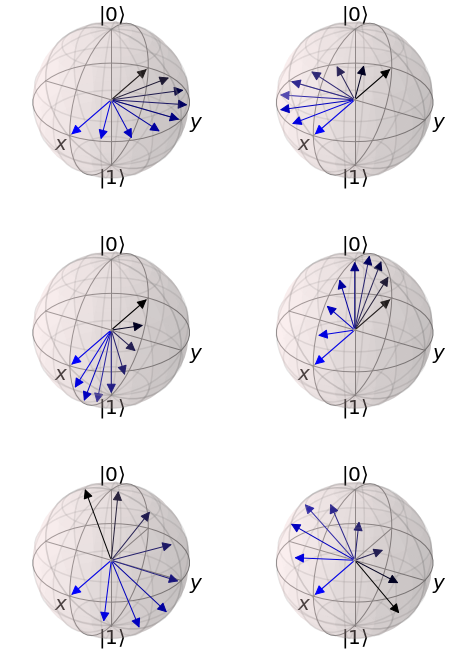}
	\caption{Various rotations of the vector $(1,0,0)$ graphically compared}
	\label{fig:all_rotations}
\end{figure}

The plots, which result from Listing~\ref{lst:album_rotate_su2_euler} are depicted by Figure~\ref{fig:all_rotations}.

Figure \ref{fig:all_rotations} above shows various rotations of the vector \((1, 0, 0)\) on the Bloch sphere. Each block sphere represents different sets of rotation operations applied to the vector, and the visual comparison helps to illustrate the behaviour of quantum states under these rotations. The rotations change the vector’s orientation on the sphere, revealing the effects of quantum gate operations.

\begin{listing}[H]
\begin{minted}[bgcolor=bg, frame=leftline, linenos, texcl=false, framesep=2mm, breaklines, tabsize=2]{python}
def rn_su2(vec, rot_angle, n):
    spherical_vec = cartesian_to_spherical(vec)
    θ = spherical_vec[1]
    φ = spherical_vec[2]
    
    sx = msigma(1)
    sy = msigma(2)
    sz = msigma(3)
    M_q = sin(θ)*cos(φ)*sx + sin(θ)*sin(φ)*sy + cos(θ)*sz
    U_n = eye(2)*cos(rot_angle/2) -I*(n[0]*sx+n[1]*sy+n[2]*sz)*sin(rot_angle/2)
    M_q_rotated = U_n*M_q*Dagger(U_n)
    return M_q_rotated
\end{minted}
\caption{Rotate a qubit around an axis as per equation~\eqref{eq:rotation_su2_axis}}
\label{lst:rotate_su2_axis}
\end{listing}

\begin{figure}[H]
	\centering
	\includegraphics[clip, trim=0cm 0cm 0cm 0cm, width=0.6\textwidth]{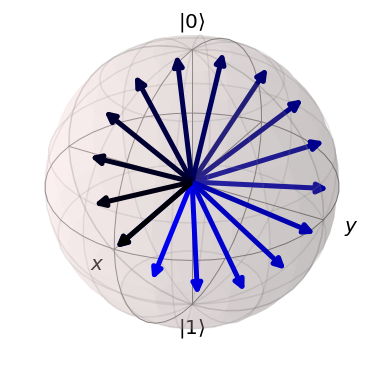}
	\caption{Plot of rotating the vector $(1,0,0)$ around the axis $(\nicefrac{1}{\sqrt{3}},\nicefrac{1}{\sqrt{3}},\nicefrac{1}{\sqrt{3}})$}
	\label{fig:axial_rotation}
\end{figure}

Rotating a qubit $\vec{q}=(1,0,0)$ around the axis $\hat n=(\nicefrac{1}{\sqrt{3}},\nicefrac{1}{\sqrt{3}},\nicefrac{1}{\sqrt{3}})$ 16 times repeatedly by an angle $\theta=\nicefrac{\pi}{8}$ leads to the resulting plots depicted by Figure~\ref{fig:axial_rotation}.

\section{Euler matrix-based rotations using Python}
\label{appx:impl-rotations-euler-matrix}

Let us refer to the plots given in Figure~\ref{fig:all_rotations} showing several rotations of the vector $(1,0,0)$ around different Euler angles. The same plots can be generated by the following Listing~\ref{lst:album_rotate_euler_matrix} which rotates the vector $(1,0,0)$ by utilizing the Euler matrix instead of a $SU(2)$ matrix.

\begin{listing}[H]
\begin{minted}[bgcolor=bg, frame=leftline, linenos, texcl=false, framesep=2mm, breaklines, tabsize=2]{python}
from sympy import rot_axis1
from sympy import rot_axis2
from sympy import rot_axis3

fig, ax = plt.subplots(figsize = [8, 12], nrows=3, ncols=2)
fig.patch.set_facecolor('white')
[axis.set_axis_off() for axis in ax.ravel()]

rotations = [[0, 0, pi/8], [0, 0, -pi/8], [0, pi/8, 0], [0, -pi/8, 0], [0, pi/8, pi/8], [0, -pi/8, -pi/8]]
start_vec = [1, 0, 0]
num_iterations = 8
for m, rotation in enumerate(rotations):
    ax = fig.add_subplot(320+(m+1), axes_class = Axes3D)
    rot_x = rotation[0]
    rot_y = rotation[1]
    rot_z = rotation[2]
    rot_mat = rot_axis3(rot_z)*rot_axis2(rot_y)*rot_axis1(rot_x)
    _bloch = Bloch(axes=ax)
    _bloch.vector_color = get_gradient_colors([0, 0, 1], num_iterations)
    _bloch.vector_width = 1
    sv = []
    vec = Matrix(start_vec).T
    sv.append(np.array(vec).astype(np.float64)[0])
    for i in range(num_iterations):
        vec = N(vec * rot_mat)
        sv.append(np.array(vec).astype(np.float64)[0])

    _bloch.add_vectors(sv)
    _bloch.render()
\end{minted}
\caption{Various rotations of the vector $(1,0,0)$ around different Euler angles using Euler matrix}
\label{lst:album_rotate_euler_matrix}
\end{listing}

\section{\texorpdfstring{Implementing $SU(2)$ based error propagation using Python}{Implementing SU(2) based error propagation using Python}}
\label{appx:impl-error-propagation-su2-python}

The Python code for error propagation analysis based on $SU(2)$ rotations is provided in Listing~\ref{lst:error_propagation_su2_python}, which is part of the Notebook \href{https://github.com/Sultanow/quantum/blob/main/bloch_sphere/rotate_su2_qiskit_eldar-sultanow.ipynb}{rotate\_su2\_qiskit\_eldar-sultanow.ipynb} and uses the functions described in Appendix~\ref{appx:impl-rotations-su2}.

\begin{listing}[H]
\begin{minted}[bgcolor=bg, frame=leftline, linenos, texcl=false, framesep=2mm, breaklines, tabsize=2]{python}
rot_x = pi/100
rot_y = pi/100
rot_z = pi/100
num_iterations = 200
x = np.arange(0, num_iterations, 1, dtype=int)
start_vec = [1, 0, 0]
err = 0.2
vec = start_vec
vec_err = qubitmatrix_to_cartesian(rn_su2_euler(start_vec, 0, err, 0))
φ_error_propagation_vec = np.zeros(shape=(num_iterations))
θ_error_propagation_vec = np.zeros(shape=(num_iterations))

for i in range(num_iterations):
    spherical = cartesian_to_spherical(vec)
    spherical_err = cartesian_to_spherical(vec_err)
    (θ_rotated, φ_rotated) = (spherical[1], spherical[2])
    (θ_rotated_err, φ_rotated_err) = (spherical_err[1], spherical_err[2])
    d_θ = abs(θ_rotated_err - θ_rotated)
    d_φ = abs(φ_rotated_err - φ_rotated)
    θ_error_propagation_vec[i] = min(d_θ, 2*pi-d_θ)
    φ_error_propagation_vec[i] = min(d_φ, 2*pi-d_φ)

    M_q_rotated = rn_su2_euler(vec, rot_x, rot_y, rot_z)
    M_q_rotated_err = rn_su2_euler(vec_err, rot_x, rot_y, rot_z)
    vec = qubitmatrix_to_cartesian(M_q_rotated)
    vec_err = qubitmatrix_to_cartesian(M_q_rotated_err)

plt.plot(x, φ_error_propagation_vec, θ_error_propagation_vec)
plt.show()
\end{minted}
\caption{Calculate error propagation using $SU(2)$ rotations (Python)}
\label{lst:error_propagation_su2_python}
\end{listing}

\begin{figure}[H]
	\centering
	\includegraphics[clip, trim=0cm 0cm 0cm 0cm, width=0.6\textwidth]{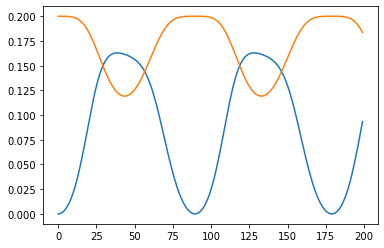}
	\caption{Curves of the error propagation (azimuth/blue and elevation/orange) due to 200 qubit rotations by $\nicefrac{\pi}{100}$ drawn by Matplotlib via Listing~\ref{lst:error_propagation_su2_python}}
	\label{fig:error_propagation_python}
\end{figure}

The resulting plot generated by Listing~\ref{lst:error_propagation_su2_python} is shown in Figure~\ref{fig:error_propagation_python}.

\section{\texorpdfstring{Implementing $SU(2)$ based error propagation using Mathematica}{Implementing SU(2) based error propagation using Mathematica}}
\label{appx:impl-error-propagation-su2-mathematica}

This section contains the $SU(2)$ rotation-based implementation of error propagation analysis using Mathematica. Listing~\ref{lst:error_propagation_su2_mathematica} contains a Mathematica Notebook for calculating and visualizing the error propagation due to 200 qubit rotations by $\nicefrac{\pi}{100}$ (see entire Notebook \href{https://github.com/Sultanow/quantum/blob/main/bloch_sphere/mathematica/errorPropagation.nb}{errorPropagation.nb} on Github). The resulting plot is shown in Figure~\ref{fig:error_propagation}.

\begin{listing}[H]
\begin{minted}[bgcolor=bg, frame=leftline, linenos, texcl=false, framesep=2mm, breaklines, tabsize=2]{mathematica}
qubitmatrixToCartesian[Mq_] := (
   q1 = Re[(Mq[[1, 2]] + Mq[[2, 1]])/2];
   q2 = Re[(Mq[[2, 1]] - Mq[[1, 2]])/(2*I)];
   q3 = Re[Mq[[1, 1]]];
   Return[{q1, q2, q3}];
   );
rnSU2euler[vec_, rx_, ry_, rz_] := (
   sphericalVec = ToSphericalCoordinates[vec];
   θ = sphericalVec[[2]];
   φ = sphericalVec[[3]];
   sx = PauliMatrix[1];
   sy = PauliMatrix[2];
   sz = PauliMatrix[3];
   Mq = Sin[θ]*Cos[φ]*sx + Sin[θ]*Sin[φ]*sy + 
     Cos[θ]*sz;
   Un = {{Exp[-I*(rx + rz)/2]*Cos[ry/2], -Exp[-I*(rx - rz)/2]*
       Sin[ry/2]}, {Exp[I*(rx - rz)/2]*Sin[ry/2], 
      Exp[I*(rx + rz)/2]*Cos[ry/2]}};
   Return [Un . Mq . ConjugateTranspose[Un]];
   );
rotateVector[vec_, rx_, ry_, rz_] := (
   Return[qubitmatrixToCartesian[rnSU2euler[vec, rx, ry, rz]]];
   );
subtractAngles[a1_, a2_] := (
   d = RealAbs[a1 - a2];
   Return[Min[d, 2*Pi - d]];
   );
errorPropagation[n_, vec_, vecError_, rx_, ry_, rz_] := (
   v = NestList[rotateVector[#, rx, ry, rz] &, N@vec, n];
   vErr = NestList[rotateVector[#, rx, ry, rz] &, N@vecError, n];
   spherical = Map[ToSphericalCoordinates, v];
   sphericalErr = Map[ToSphericalCoordinates, vErr];
   {MapThread[subtractAngles, {sphericalErr[[All, 3]], spherical[[All, 3]]}], 
    MapThread[subtractAngles, {sphericalErr[[All, 2]], spherical[[All, 2]]}]}
   );
ListLinePlot[
 errorPropagation[200, {1, 0, 0}, 
  N[rotateVector[{1, 0, 0}, 0, 0.2, 0]], Pi/100, Pi/100, Pi/100]]
\end{minted}
\caption{Calculate and plot error propagation using $SU(2)$ rotations (Mathematica)}
\label{lst:error_propagation_su2_mathematica}
\end{listing}

\section{Implementing Euler matrix-based error propagation using Mathematica}
\label{appx:impl-error-propagation-euler-matrix-mathematica}

This section contains the Euler matrix-based implementation of error propagation analysis using Mathematica. Listing~\ref{lst:error_propagation_euler_matrix_mathematica} contains a Mathematica Notebook for calculating and visualizing the error propagation due to 200 qubit rotations by $\nicefrac{\pi}{100}$ (see entire Notebook \href{https://github.com/Sultanow/quantum/blob/main/bloch_sphere/mathematica/errorPropagation.nb}{errorPropagation.nb} on Github). The resulting plot looks exactly the same as the one generated by Listing~\ref{lst:error_propagation_su2_mathematica} shown in Figure~\ref{fig:error_propagation}.

\begin{listing}[H]
\begin{minted}[bgcolor=bg, frame=leftline, linenos, texcl=false, framesep=2mm, breaklines, tabsize=2]{mathematica}
angularError[vec_, vecError_, rx_, ry_, rz_, t_, i_] := (
   sphericalVec = 
    ToSphericalCoordinates[
     vec . MatrixPower[N[EulerMatrix[{rx, ry, rz}]], t]];
   sphericalVecError = 
    ToSphericalCoordinates[
     vecError . MatrixPower[N[EulerMatrix[{rx, ry, rz}]], t]];
   Return[subtractAngles[sphericalVec[[i]], sphericalVecError[[i]]]];
   );
elevationErrorPlot = 
  Plot[angularError[{1, 0, 0}, 
    N[{1, 0, 0} . EulerMatrix[{0, 0.2, 0}]], Pi/100, Pi/100, Pi/100, 
    t, 2], {t, 0, 200}, PlotStyle -> Orange];
azimuthalErrorPlot = 
  Plot[angularError[{1, 0, 0}, 
    N[{1, 0, 0} . EulerMatrix[{0, 0.2, 0}]], Pi/100, Pi/100, Pi/100, 
    t, 3], {t, 0, 200}];
Show[azimuthalErrorPlot, elevationErrorPlot, PlotRange -> All]
\end{minted}
\caption{Calculate and plot error propagation using Euler matrix (Mathematica)}
\label{lst:error_propagation_euler_matrix_mathematica}
\end{listing}

\begin{listing}[H]
\begin{minted}[bgcolor=bg, frame=leftline, linenos, texcl=false, framesep=2mm, breaklines, tabsize=2]{mathematica}
SP[t_] := {{Cos[\[Sqrt]5 t], -((2 Sin[\[Sqrt]5 t])/(\[Sqrt]5)), 
    Sin[\[Sqrt]5 t]/(\[Sqrt]5)}, {(2 Sin[\[Sqrt]5 t])/(\[Sqrt]5), (1/5) (1 + 4 Cos[\[Sqrt]5 t]), -(2/5) (-1 + 
    Cos[\[Sqrt]5 t])}, {-(Sin[\[Sqrt]5 t]/(\[Sqrt]5)), -(2/5) (-1 + Cos[\[Sqrt]5 t]), (1/5) (4 + Cos[\[Sqrt]5 t])}};
angularErrorSimplified[vec_, vecError_, t_, i_] := (
   sphericalVec = ToSphericalCoordinates[vec . SP[t]];
   sphericalVecError = ToSphericalCoordinates[vecError . SP[t]];
   Return[subtractAngles[sphericalVec[[i]], sphericalVecError[[i]]]];
   );
elevationErrorPlotSimplified = 
  Plot[angularErrorSimplified[{1, 0, 0}, 
    N[{1, 0, 0} . EulerMatrix[{0, 0.2, 0}]], t, 2], {t, 0, 2 Pi}, 
   PlotStyle -> Orange];
azimuthalErrorPlotSimplified = 
  Plot[angularErrorSimplified[{1, 0, 0}, 
    N[{1, 0, 0} . EulerMatrix[{0, 0.2, 0}]], t, 3], {t, 0, 2 Pi}];
Show[azimuthalErrorPlotSimplified, elevationErrorPlotSimplified,  PlotRange -> All]
\end{minted}
\caption{Calculate and plot error propagation using the limit of Euler matrix power (Mathematica)}
\label{lst:error_propagation_limit_euler_matrix_power_mathematica}
\end{listing}
\newpage
\section{Poisoning Algorithms}
\label{appx:Poisoning}
For simplicity, we present the QNN poisoning algorithm\cite{kundu2024adversarialpoisoningattackquantum} here. This algorithm is not used in the SVM or QSVM experiments in the main text but is included here for reference, as it inspired our new QUID-based poisoning methods.

The recursive versions of Algorithms ~\ref{alg:poision_svm} and ~\ref{alg:poision_qsvm} provide an alternative way to demonstrate how induction can be used to prove QSVM resiliency against poisoning. Here, we present the recursive versions and their equivalency with normal version.\\ 

\begin{algorithm}[H]
\caption{Original QUID’s Label Poisoning Procedure for QNN}
\label{alg:poision_qnn}

\textbf{Require:} Training data $\mathcal{D}_{\text{tr}} = \{(x_i, y_i)\}_{i=1}^n$, Poison ratio $\epsilon$, Encoding circuit $\phi$, Distance metric $d(\cdot, \cdot)$ for density matrices.\
\textbf{Ensure:} Poisoned dataset with modified labels.
\BlankLine
Split $\mathcal{D}_{\text{tr}}$ into clean set $\mathcal{D}_c$ and poison set $\mathcal{D}_p$ with ratio $\epsilon$\;
$\mathcal{C} \gets \text{unique}(\{y_i \mid (x_i, y_i) \in \mathcal{D}_{\text{tr}}\})$ \tcp{Set of unique classes.}
$\rho_c \gets \{\phi(x) \mid (x, y) \in \mathcal{D}_c\}$ \tcp{Encoded clean states.}
$\rho_p \gets \{\phi(x) \mid (x, y) \in \mathcal{D}_p\}$ \tcp{Encoded poison states.}
\For{$\rho_i \in \rho_p$}{
    $D_{\text{cls}} \gets \{\}$ \tcp{Initialize dictionary for class-wise distances.}
    \For{$c \in \mathcal{C}$}{
        $\rho_c^{(c)} \gets \{\rho \mid \rho \in \rho_c, y = c\}$ \tcp{States of class $c$.}
        $D_{\text{cls}}[c] \gets \frac{1}{|\rho_c^{(c)}|} \sum_{\rho \in \rho_c^{(c)}} d(\rho_i, \rho)$\;
    }
    $y_i^{\text{new}} \gets \arg\max_{c \in \mathcal{C}} D_{\text{cls}}[c]$ \tcp{Assign class with maximum distance.}
}
\Return{$\mathcal{D}_c \cup \{(x_i, y_i^{\text{new}}) \mid i \in \mathcal{D}_p\}$}\;
\label{alg:alg1}
\end{algorithm}

\begin{algorithm}[H]
\caption{Recursive QUID-style Label Poisoning for Classical SVM}
\label{alg:poision_reqsvm}

\textbf{Require:} Training data $\mathcal{D}_{\text{tr}} = \{(x_i, y_i)\}_{i=1}^n$, Poison ratio $\epsilon$, Kernel function $k(x, x')$, Distance metric $d(\cdot, \cdot)$.\
\textbf{Ensure:} Poisoned dataset with modified labels.
\BlankLine

Split $\mathcal{D}_{\text{tr}}$ into clean set $\mathcal{D}_c$ and poison set $\mathcal{D}_p$ with ratio $\epsilon$\;
$\mathcal{C} \gets \text{unique}(\{y_i \mid (x_i, y_i) \in \mathcal{D}_{\text{tr}}\})$ \tcp{Set of unique classes.}
$\Phi_c \gets \{k(x, x') \mid (x, y) \in \mathcal{D}_c\}$ \tcp{Kernel-induced clean feature space.}
$\Phi_p \gets \{k(x, x') \mid (x, y) \in \mathcal{D}_p\}$ \tcp{Kernel-induced poison feature space.}

\If{$|\Phi_p| \le 1$}{
    \tcp{Base case: no or single point to poison}
    \Return $\mathcal{D}_{\text{tr}}$ \;
}
\Else{
    \tcp{Recursive case}
    Let $\phi_{\text{curr}}$ be the first element in $\Phi_p$ (corresponding to $(x_{\text{curr}}, y_{\text{curr}})$)\;
    $D_{\text{cls}} \gets \{\}$ \tcp{Initialize dictionary for class-wise distances.}
    \For{$c \in \mathcal{C}$}{
        $\Phi_c^{(c)} \gets \{\phi \mid \phi \in \Phi_c, y = c\}$ \tcp{Features of class $c$.}
        $D_{\text{cls}}[c] \gets \frac{1}{|\Phi_c^{(c)}|} \sum_{\phi \in \Phi_c^{(c)}} d(\phi_{\text{curr}}, \phi)$\;
    }
    $y_{\text{curr}}^{\text{new}} \gets \arg\max_{c \in \mathcal{C}} D_{\text{cls}}[c]$ \tcp{Assign class with maximum distance.}

    Update label of $(x_{\text{curr}}, y_{\text{curr}})$ in $\mathcal{D}_p$ to $y_{\text{curr}}^{\text{new}}$\;
    Remove $\phi_{\text{curr}}$ from $\Phi_p$ and the corresponding sample from $\mathcal{D}_p$\;

    $\mathcal{D}_{\text{tr}}' \gets \mathcal{D}_c \cup \mathcal{D}_p$ \tcp{Updated dataset}
    
    \tcp{Recursively poison the remaining points}
    $\mathcal{D}_{\text{tr}}'' \gets \text{RecursiveQUIDSVM}(\mathcal{D}_{\text{tr}}', \epsilon, k, d)$\;

    \Return $\mathcal{D}_{\text{tr}}''$ \;
}
\end{algorithm}

\medskip
\textbf{Theorem (Equivalence).}
Let \(\text{QUIDSVM}\) denote the original, non-recursive QUID-style Label Poisoning for Classical SVM (Algorithm~\ref{alg:poision_svm}). Let \(\text{RecursiveQUIDSVM}\) be its recursive version, as defined in Algorithm~4. For any training dataset \(\mathcal{D}_{\text{tr}}\), poison ratio \(\epsilon\), kernel function \(k(\cdot,\cdot)\), and distance metric \(d(\cdot,\cdot)\), \emph{both} algorithms yield the \emph{same} final set of label assignments. In other words,
\[
   \text{QUIDSVM}(\mathcal{D}_{\text{tr}}, \epsilon, k, d) 
   \;=\;
   \text{RecursiveQUIDSVM}(\mathcal{D}_{\text{tr}}, \epsilon, k, d).
\]

\textbf{Proof.} 
We prove the statement by \emph{induction on the size of the poison set}.

\medskip
\noindent
\emph{Notation.} For a dataset \(\mathcal{D}_{\text{tr}}\), let 
\[
   \mathcal{D}_c \subseteq \mathcal{D}_{\text{tr}}, 
   \quad 
   \mathcal{D}_p \subseteq \mathcal{D}_{\text{tr}}
\]
be the clean and poison partitions, respectively, after splitting with ratio \(\epsilon\). Define 
\(\Phi_c = \{\,k(x,x')\mid(x,y)\in\mathcal{D}_c\}\) and 
\(\Phi_p = \{\,k(x,x')\mid(x,y)\in\mathcal{D}_p\}\) 
as in Algorithm~\ref{alg:poision_svm}.\\

\begin{description}
  \item[Base Case] (\(|\Phi_p|\le 1\)). 
    If there are zero or one points in \(\Phi_p\), the poisoning procedure performs \emph{at most} one label flip. In both the non-recursive and recursive versions, that single flip (or no flip) is carried out identically. Hence, the result is trivially the same.

  \item[Inductive Step] 
    Assume the claim holds for any dataset whose poison set \(\Phi_p\) has size \(k\). 
    Now consider a dataset with \(|\Phi_p|=k+1\). 
    \begin{enumerate}
        \item In the \textbf{non-recursive} Algorithm~~\ref{alg:poision_svm} (\texttt{QUIDSVM}), we have a loop:
        \begin{align*}
           \textbf{for} \quad \phi_i \in \Phi_p \quad \{\dots\}.
        \end{align*}
        That loop processes each \(\phi_i\) in turn, computing the class-wise distances 
        \(\displaystyle 
            D_{\text{cls}}[c]
        \)
        and then assigning the new label \(\arg\max_c\, D_{\text{cls}}[c].\)

        \item In the \textbf{recursive} version (\texttt{RecursiveQUIDSVM}), we pick the \emph{first} point \(\phi_{\text{curr}}\) from \(\Phi_p\), perform the same distance-based label assignment, then remove \(\phi_{\text{curr}}\) from \(\Phi_p\). This reduces the poison set to size \(k\). By the \emph{inductive hypothesis}, calling
        \[
          \text{RecursiveQUIDSVM}(\mathcal{D}_{\text{tr}}', \epsilon, k, d)
        \]
        on the remaining \(k\) poison points yields exactly the same final set of label assignments as the non-recursive procedure would, once it had moved on from \(\phi_{\text{curr}}\).

        Since the update step for \(\phi_{\text{curr}}\) is also identical in both algorithms (same \(\arg\max\) rule, same distances, etc.), the entire sequence of label flips (on all \(k+1\) points) ends up the same.
    \end{enumerate}
    
    Therefore, by induction, both algorithms produce the same final labelling whenever \(|\Phi_p|=k+1\).
\end{description}

Since both the base case and the inductive step are verified, the \emph{Equivalence Theorem} holds for all possible sizes of the poison set. 
\(\quad\Box\)

\newpage
\begin{algorithm}[H]
\caption{Recursive QUID-style Label Poisoning for QSVM}
\label{alg:poision_reqqsvm}

\textbf{Require:} Training data $\mathcal{D}_{\text{tr}} = \{(x_i, y_i)\}_{i=1}^n$, Poison ratio $\epsilon$, Encoding circuit $\phi$, Distance metric $d(\cdot, \cdot)$ for density matrices.\
\textbf{Ensure:} Poisoned dataset with modified labels.
\BlankLine

Split $\mathcal{D}_{\text{tr}}$ into clean set $\mathcal{D}_c$ and poison set $\mathcal{D}_p$ with ratio $\epsilon$\;
$\mathcal{C} \gets \text{unique}(\{y_i \mid (x_i, y_i) \in \mathcal{D}_{\text{tr}}\})$ \tcp{Set of unique classes.}
$\rho_c \gets \{\phi(x) \mid (x, y) \in \mathcal{D}_c\}$ \tcp{Encoded clean states.}
$\rho_p \gets \{\phi(x) \mid (x, y) \in \mathcal{D}_p\}$ \tcp{Encoded poison states.}

\If{$|\rho_p| \le 1$}{
    \tcp{Base case: no or single point to poison}
    \Return $\mathcal{D}_{\text{tr}}$ \;
}
\Else{
    \tcp{Recursive case}
    Let $\rho_{\text{curr}}$ be the first element in $\rho_p$ (corresponding to $(x_{\text{curr}}, y_{\text{curr}})$)\;
    $D_{\text{cls}} \gets \{\}$ \tcp{Initialize dictionary for class-wise distances.}
    \For{$c \in \mathcal{C}$}{
        $\rho_c^{(c)} \gets \{\rho \mid \rho \in \rho_c, y = c\}$ \tcp{States of class $c$.}
        $D_{\text{cls}}[c] \gets \frac{1}{|\rho_c^{(c)}|} \sum_{\rho \in \rho_c^{(c)}} d(\rho_{\text{curr}}, \rho)$\;
    }
    $y_{\text{curr}}^{\text{new}} \gets \arg\max_{c \in \mathcal{C}} D_{\text{cls}}[c]$ \tcp{Assign class with maximum distance.}

    Update label of $(x_{\text{curr}}, y_{\text{curr}})$ in $\mathcal{D}_p$ to $y_{\text{curr}}^{\text{new}}$\;
    Remove $\rho_{\text{curr}}$ from $\rho_p$ and the corresponding sample from $\mathcal{D}_p$\;

    $\mathcal{D}_{\text{tr}}' \gets \mathcal{D}_c \cup \mathcal{D}_p$ \tcp{Updated dataset}

    \tcp{Recursively poison the remaining points}
    $\mathcal{D}_{\text{tr}}'' \gets \text{RecursiveQUIDQSVM}(\mathcal{D}_{\text{tr}}', \epsilon, \phi, d)$\;
    
    \Return $\mathcal{D}_{\text{tr}}''$ \;
}
\end{algorithm}

\medskip
\textbf{Theorem (Equivalence for QSVM).}
Let \(\text{QUIDQSVM}\) be the original, non-recursive QUID-style Label Poisoning for QSVM (Algorithm~\ref{alg:poision_qsvm}). 
Let \(\text{RecursiveQUIDQSVM}\) be its recursive version, as defined in Algorithm~5. 
For any training dataset \(\mathcal{D}_{\text{tr}}\), poison ratio \(\epsilon\), encoding circuit \(\phi\), and distance metric \(d(\cdot,\cdot)\) on density matrices, both algorithms yield the \emph{same} final poisoned dataset. 
Symbolically,
\[
   \text{QUIDQSVM}(\mathcal{D}_{\text{tr}}, \epsilon, \phi, d) 
   \;=\;
   \text{RecursiveQUIDQSVM}(\mathcal{D}_{\text{tr}}, \epsilon, \phi, d).
\]

\textbf{Proof.}
We proceed by \emph{induction on the size} of the poison set.

\medskip
\noindent
\emph{Notation.}
From the training set \(\mathcal{D}_{\text{tr}}\), let 
\[
   \mathcal{D}_c \subseteq \mathcal{D}_{\text{tr}}, 
   \quad 
   \mathcal{D}_p \subseteq \mathcal{D}_{\text{tr}}
\]
be the clean and poison subsets after splitting with ratio \(\epsilon\). 
Define 
\[
   \rho_c \;=\; \{\,\phi(x) \,\mid\, (x,y)\in \mathcal{D}_c\},
   \quad
   \rho_p \;=\; \{\,\phi(x) \,\mid\, (x,y)\in \mathcal{D}_p\},
\]
as in Algorithm~2.\\

\begin{description}
  \item[Base Case (\(|\rho_p|\le 1\))] 
    If \(\rho_p\) has size 0 or 1, then there are at most one or zero label reassignments to perform. 
    Both the \(\text{QUIDQSVM}\) and the \(\text{RecursiveQUIDQSVM}\) do the same update (or no update) in this scenario, producing the same final dataset. 
    Thus, the claim holds trivially.

  \item[Inductive Step] 
    Suppose that for any dataset whose poison set has size \(k\), both algorithms return identical label assignments. 
    Consider a dataset for which \(|\rho_p| = k+1\). 

    \begin{enumerate}
      \item \textbf{Non-Recursive Algorithm~\ref{alg:poision_qsvm}(\texttt{QUIDQSVM}).} 
      It iterates over all states in \(\rho_p\) (say \(\rho_1,\rho_2,\dots\)), each time:
      \[
         D_{\text{cls}}[c] 
         \;=\;
         \frac{1}{|\rho_c^{(c)}|} 
         \sum_{\rho \in \rho_c^{(c)}} 
            d(\rho_i, \rho)
         \quad \text{for each class } c,
         \quad
         y_i^{\text{new}} 
         \;=\;
         \arg\max_{c}\,D_{\text{cls}}[c].
      \]
      Labels are updated one by one in a for-loop.

      \item \textbf{Recursive Algorithm~5 (\texttt{RecursiveQUIDQSVM}).} 
      It picks the \emph{first} state \(\rho_{\text{curr}}\in \rho_p\), computes the same distances 
      \(\{D_{\text{cls}}[c]\}\), 
      and flips its label using the same \(\arg\max\) rule. 
      Then it \emph{removes} \(\rho_{\text{curr}}\) from \(\rho_p\), leaving a poison set of size \(k\). 
      By the inductive hypothesis, the recursive call on that reduced set yields the same final labelling as the non-recursive version would do for the \emph{remaining} \(k\) points (after it, too, finishes flipping the label of \(\rho_{\text{curr}}\) and moves on).

      Because the step for \(\rho_{\text{curr}}\) is identical in both algorithms (the same distance calculations and \(\arg\max\)), the entire sequence of \((k+1)\) label flips is the same overall.
    \end{enumerate}

    Hence by induction, both algorithms produce an identical final labeling whenever \(|\rho_p|=k+1\).
\end{description}

Since both the base case and the inductive step have been shown, \emph{the two algorithms are equivalent for all sizes} of the poison set. 
\(\quad\Box\)

\newpage
\section{Glossary}
\label{appx:glossary}

\begin{table}[H]
\centering
\footnotesize 
\setlength{\tabcolsep}{0.8em} 
\setlength\extrarowheight{2pt} 
\begin{tabular}{|p{3.5cm}|p{9cm}|}
\hline
\textbf{Term} & \textbf{Definition}\\
\hline
Azimuthal Angle ($\varphi_{az}$) &
In the Bloch sphere representation of a qubit, the azimuthal angle $\varphi_{az}$ is the angle between the projection of the vector $\vec{q}$ onto the $xy$-plane and the positive $x$-axis. It ranges from $0$ to $2\pi$. It is used to express the qubit in spherical coordinates. \\
\hline
Elevation Angle ($\theta_{el}$) &
In the Bloch sphere representation of a qubit, the elevation angle $\theta_{el}$ (also known as the polar angle) is the angle between the vector $\vec{q}$ and the positive $z$-axis. It ranges from $0$ to $\pi$. It is used to express the qubit in spherical coordinates. \\
\hline
Data Poisoning &
An attack on machine learning models where an adversary manipulates the training data to influence the behaviour of the trained model. In the context of the paper, it refers to injecting errors into quantum machine learning models to study how errors propagate. \\
\hline
Error Propagation &
The process by which errors are introduced at one point in a computational process affects subsequent computations. The paper investigates how errors propagate in quantum machine learning, particularly in qubit rotations. \\
\hline
Euler Matrix &
A rotation matrix constructed from Euler angles $(\varphi, \theta, \psi)$, representing a rotation in three-dimensional space. In the paper, it is used to rotate qubits in their Cartesian representation. \\
\hline
Generator of Rotation ($J$) &
An operator or matrix that generates infinitesimal rotations. In the paper, $J$ is used to express finite rotations as exponentials of $J$, considering infinitesimal rotations successively. \\
\hline
Infinitesimal Rotation \cite{ryparova2019infinitesimal} &
A rotation by an infinitesimally small angle. In the paper, infinitesimal rotations are used to derive expressions for finite rotations by taking the limit as the number of rotations approaches infinity. \\
\hline
Norm-Preserving &
A property of a transformation that preserves the norm (length) of vectors it acts upon. In quantum computing, unitary operators are norm-preserving transformations. \\
\hline
Periodicity &
The quality of a function or process that repeats at regular intervals. In the paper, error propagation functions exhibit periodic behaviour due to the properties of qubit rotations on the Bloch sphere. \\
\hline
\end{tabular}
\end{table}

\begin{table}[H]
\centering
\footnotesize 
\setlength{\tabcolsep}{0.8em} 
\setlength\extrarowheight{2pt} 
\begin{tabular}{|p{3.5cm}|p{9cm}|}
\hline
\textbf{Term} & \textbf{Definition}\\
\hline
Special Unitary Group $SU(2)$ \cite{sethuraman2006note} &
The group of $2 \times 2$ unitary matrices with determinant 1. In the paper, $SU(2)$ matrices are used to describe rotations of qubits on the Bloch sphere. \\
\hline
Traceless Matrix \cite{garcia2020differentiable} &
A square matrix whose trace (the sum of the diagonal elements) is zero. \\
\hline
Unitary Operators &
Operators that preserve inner products in a Hilbert space satisfy $U^\dagger U = U U^\dagger = I$, where $U^\dagger$ is the conjugate transpose of $U$ and $I$ is the identity operator. \\
\hline
\end{tabular}
\end{table}

\newpage
\vspace{1em}
\bibliographystyle{unsrt}
\bibliography{main}

\begin{thebibliography}{10}

\bibitem{yerlikaya2022data}
Fahri~An{\i}l Yerlikaya and {\c{S}}erif Bahtiyar.
\newblock Data poisoning attacks against machine learning algorithms.
\newblock {\em Expert Systems with Applications}, 208:118101, 2022.

\bibitem{lu2020quantum}
Sirui Lu, Lu-Ming Duan, and Dong-Ling Deng.
\newblock Quantum adversarial machine learning.
\newblock {\em Physical Review Research}, 2(3):033212, 2020.

\bibitem{biamonte2017quantum}
Jacob Biamonte, Peter Wittek, Nicola Pancotti, Patrick Rebentrost, Nathan
  Wiebe, and Seth Lloyd.
\newblock Quantum machine learning.
\newblock {\em Nature}, 549(7671):195--202, 2017.

\bibitem{abohashima2020classification}
Zainab Abohashima, Mohamed Elhosen, Essam~H Houssein, and Waleed~M Mohamed.
\newblock Classification with quantum machine learning: A survey.
\newblock {\em arXiv preprint arXiv:2006.12270}, 2020.

\bibitem{goyal2016geometry}
Sandeep~K Goyal, B~Neethi Simon, Rajeev Singh, and Sudhavathani Simon.
\newblock Geometry of the generalized bloch sphere for qutrits.
\newblock {\em Journal of Physics A: Mathematical and Theoretical},
  49(16):165203, 2016.

\bibitem{tilma2002generalized}
Todd Tilma and ECG Sudarshan.
\newblock Generalized euler angle parametrization for su (n).
\newblock {\em Journal of Physics A: Mathematical and General}, 35(48):10467,
  2002.

\bibitem{wie2020two}
Chu-Ryang Wie.
\newblock Two-qubit bloch sphere.
\newblock {\em Physics}, 2(3):383--396, 2020.

\bibitem{biggio2013poisoningattackssupportvector}
Battista Biggio, Blaine Nelson, and Pavel Laskov.
\newblock Poisoning attacks against support vector machines, 2013.

\bibitem{franco2024predominantaspectssecurityquantum}
Nicola Franco, Alona Sakhnenko, Leon Stolpmann, Daniel Thuerck, Fabian Petsch,
  Annika Rüll, and Jeanette~Miriam Lorenz.
\newblock Predominant aspects on security for quantum machine learning:
  Literature review, 2024.

\bibitem{wendlinger2024comparativeanalysisadversarialrobustness}
Maximilian Wendlinger, Kilian Tscharke, and Pascal Debus.
\newblock A comparative analysis of adversarial robustness for quantum and
  classical machine learning models, 2024.

\bibitem{kundu2024securityconcernsquantummachine}
Satwik Kundu and Swaroop Ghosh.
\newblock Security concerns in quantum machine learning as a service, 2024.

\bibitem{li2024computablemodelindependentboundsadversarial}
Bacui Li, Tansu Alpcan, Chandra Thapa, and Udaya Parampalli.
\newblock Computable model-independent bounds for adversarial quantum machine
  learning, 2024.

\bibitem{kundu2024adversarialpoisoningattackquantum}
Satwik Kundu and Swaroop Ghosh.
\newblock Adversarial poisoning attack on quantum machine learning models,
  2024.

\bibitem{10688912}
Sijia Yu and Yifan Zhou.
\newblock Quantum adversarial machine learning for robust power system
  stability assessment.
\newblock In {\em 2024 IEEE Power \& Energy Society General Meeting (PESGM)},
  pages 1--5, 2024.

\bibitem{reers2024comparative}
Volker Reers and Marc Mau{\ss}ner.
\newblock Comparative analysis of vulnerabilities in classical and quantum
  machine learning.
\newblock In {\em INFORMATIK 2024}, pages 555--571. Gesellschaft f{\"u}r
  Informatik eV, 2024.

\bibitem{MathWorks2024}
MathWorks.
\newblock Radar target classification using machine learning and deep learning.
\newblock
  \url{https://uk.mathworks.com/help/radar/ug/radar-target-classification-using-machine-learning-and-deep-learning.html}.
\newblock Accessed 2024-12-28.

\bibitem{adversarial_robustness_toolbox}
IBM Trusted~AI Team.
\newblock Adversarial robustness toolbox (art).
\newblock \url{https://github.com/Trusted-AI/adversarial-robustness-toolbox},
  2024.
\newblock Accessed: 2024-12-28.

\bibitem{wie2014bloch}
Chu-Ryang Wie.
\newblock Bloch sphere model for two-qubit pure states.
\newblock {\em arXiv preprint arXiv:1403.8069}, 2014.

\bibitem{Normand1980}
Jean-Marie Normand.
\newblock {\em A Lie Group, Rotations in Quantum Mechanics}.
\newblock North-Holland, 1980.

\bibitem{Yepez2013}
Jeffrey Yepez.
\newblock Lecture notes: Qubit representations and rotations.
\newblock
  \url{https://www.phys.hawaii.edu/~yepez/Spring2013/lectures/Lecture1_Qubits_Notes.pdf},
  1 2013.
\newblock Accessed 2022-06-27.

\bibitem{Arfken2013}
George~B. Arfken, Hans~J. Weber, and Frank~E. Harris.
\newblock {\em Mathematical Methods for Physicists: A Comprehensive Guide}.
\newblock Elsevier, 7 edition, 2013.

\bibitem{quintana2018euler}
Yamilet Quintana, William Ram{\'\i}rez, and Alejandro Urieles.
\newblock Euler matrices and their algebraic properties revisited.
\newblock {\em arXiv preprint arXiv:1811.01455}, 2018.

\bibitem{chruscinski2021constraints}
Dariusz Chru{\'s}ci{\'n}ski, Ryohei Fujii, Gen Kimura, and Hiromichi Ohno.
\newblock Constraints for the spectra of generators of quantum dynamical
  semigroups.
\newblock {\em Linear Algebra and its Applications}, 630:293--305, 2021.

\bibitem{Wirsching2004}
Günther~J. Wirsching.
\newblock {\em Gewöhnliche Differentialgleichungen: Eine Einführung mit
  Beispielen, Aufgaben und Musterlösungen}.
\newblock Teubner, 2006.

\bibitem{ryparova2019infinitesimal}
Lenka R{\`y}parov{\'a} and Josef Mike{\v{s}}.
\newblock Infinitesimal rotary transformation.
\newblock {\em Filomat}, 33(4):1153--1157, 2019.

\bibitem{sethuraman2006note}
B~Sethuraman and B~Sury.
\newblock A note on the special unitary group of a division algebra.
\newblock {\em Proceedings of the American Mathematical Society},
  134(2):351--354, 2006.

\bibitem{garcia2020differentiable}
M~Isabel Garc{\'\i}a-Planas and Tetiana Klymchuk.
\newblock Differentiable families of traceless matrix triples.
\newblock {\em Revista de la Real Academia de Ciencias Exactas, F{\'\i}sicas y
  Naturales. Serie A. Matem{\'a}ticas}, 114(1):11, 2020.

\end{thebibliography}



\end{document}